\documentclass[twocolumn]{aastex6}
\usepackage{apjfonts} 
\usepackage{amsmath} 
\usepackage{natbib}

\newcommand{\teff}{$T_{\rm eff}$}
 
\newcommand{\fsed}{$f_{\rm sed}$}

\newcommand{\scitarg}{WISE 0855}

\begin{document}
\title{An L Band Spectrum of the Coldest Brown Dwarf}

\author{Caroline V. Morley\altaffilmark{1,2}, Andrew J. Skemer\altaffilmark{3}, Katelyn N. Allers\altaffilmark{4}, Mark. S. Marley\altaffilmark{5}, Jacqueline K. Faherty\altaffilmark{6}, Channon Visscher\altaffilmark{7, 8}, Samuel A. Beiler\altaffilmark{4, 9}, Brittany E. Miles\altaffilmark{3}, Roxana Lupu\altaffilmark{10}, Richard S. Freedman\altaffilmark{5,11}, Jonathan J. Fortney\altaffilmark{3},  Thomas R. Geballe\altaffilmark{12}, Gordon L. Bjoraker\altaffilmark{13}  }
\date{April 2018}
\shorttitle{L Band Spectrum of WISE 0855}
\shortauthors{Morley et al.}

\altaffiltext{1}{Department of Astronomy, Harvard University, caroline.morley@cfa.harvard.edu} 
\altaffiltext{2}{Sagan Fellow} 
\altaffiltext{3}{Department of Astronomy \& Astrophysics, University of California Santa Cruz} 
\altaffiltext{4}{Department of Physics \& Astronomy, Bucknell University} 
\altaffiltext{5}{NASA Ames Research Center} 
\altaffiltext{6}{Department of Astrophysics, American Museum of Natural History} 
\altaffiltext{7}{Department of Chemistry, Dordt College} 
\altaffiltext{8}{Space Science Institute, Boulder, CO}

\altaffiltext{9}{Department of Physics, Grove City College}
\altaffiltext{10}{BAER Institute, NASA Research Park, Moffett Field, CA 94035}
\altaffiltext{11}{SETI Institute}
\altaffiltext{12}{Gemini Observatory} 
\altaffiltext{13}{NASA Goddard Space Flight Center}

\begin{abstract}

The coldest brown dwarf, WISE 0855, is the closest known planetary-mass, free-floating object and has a temperature nearly as cold as the solar system gas giants. Like Jupiter, it is predicted to have an atmosphere rich in methane, water, and ammonia, with clouds of volatile ices. WISE 0855 is faint at near-infrared wavelengths and emits almost all its energy in the mid-infrared. \citet{Skemer16} presented a spectrum of WISE 0855 from 4.5--5.1 \micron\ (M band), revealing water vapor features. Here, we present a spectrum of WISE 0855 in L band, from 3.4--4.14 \micron. We present a set of atmosphere models that include a range of compositions (metallicities and C/O ratios) and water ice clouds. Methane absorption is clearly present in the spectrum. The mid-infrared color can be better matched with a methane abundance that is depleted relative to solar abundance. We find that there is evidence for water ice clouds in the M band spectrum, and we find a lack of phosphine spectral features in both the L and M band spectra. We suggest that a deep continuum opacity source may be obscuring the near-infrared flux, possibly a deep phosphorous-bearing cloud, ammonium dihyrogen phosphate. Observations of WISE 0855 provide critical constraints for cold planetary atmospheres, bridging the temperature range between the long-studied solar system planets and accessible exoplanets. \emph{JWST} will soon revolutionize our understanding of cold brown dwarfs with high-precision spectroscopy across the infrared, allowing us to study their compositions and cloud properties, and to infer their atmospheric dynamics and formation processes. 

\end{abstract}

\keywords{planets and satellites: atmospheres, planets and satellites: gaseous planets, stars: brown dwarfs, stars: atmospheres }

\maketitle

\section{Introduction}\label{introduction}

Planetary-mass free-floating brown dwarfs serve as touchstone objects for understanding planetary atmospheres. Most of these objects likely have the bulk compositions of stars, but the physics and chemistry that govern their atmospheres are complex and closely resemble those of giant planets. 

WISE J085510.83-071442.5, hereafter WISE 0855, is unique and valuable amongst the family of known free-floating planets. It is the coldest such object discovered to date and has a temperature of $\sim$250 K, only $\sim$100 K warmer than that of Jupiter \citep{Luhman14}. WISE 0855 is one of our closest neighbors---the fourth closest system to the Solar System---with a distance measured to be 2.23 $\pm$ 0.04 parsec \citep{Luhman16}, so it provides the best opportunity to characterize a very cool substellar atmosphere. Its mass is very likely below the deuterium-burning limit: 3--10 M$_J$ assuming an age of 1--10 Gyr \citep{Luhman14}. Compared to currently known extrasolar gas giants that will be observed at high fidelity in the coming decade, WISE 0855 is the most similar object to the solar system gas giants. 

With an effective temperature of $\sim$250 K, the spectrum of WISE 0855 is expected to be dominated by water, methane, and ammonia, much like Jupiter. While hotter brown dwarfs are enshrouded in refractory iron, silicate, sulfide, and salt clouds \citep{Tsuji96, Allard01, Marley02, Burrows06, Helling08, Cushing08, Witte11, Morley12}, brown dwarfs cooler than \teff$\sim$350--375 K likely have volatile clouds of water ice \citep{Burrows03b, Morley14a}.

\subsection{Previous Photometric Observations of WISE 0855} \label{prev_photom}

A number of studies from both the ground and space have aimed to detect WISE 0855 and characterize its temperature, composition, and cloud properties. The first ground-based near-infrared observations found that WISE 0855 is very faint at these wavelengths; upper limits were placed on its flux in $z′$, $Y$, and $H$ bands \citep{Beamin14, Kopytova14, Wright14} while \citet{Faherty14b} detected WISE 0855 at 2.6$\sigma$ confidence in a deep $J$ band observation. 

\begin{figure*}[t]
\center \includegraphics[width=5in]{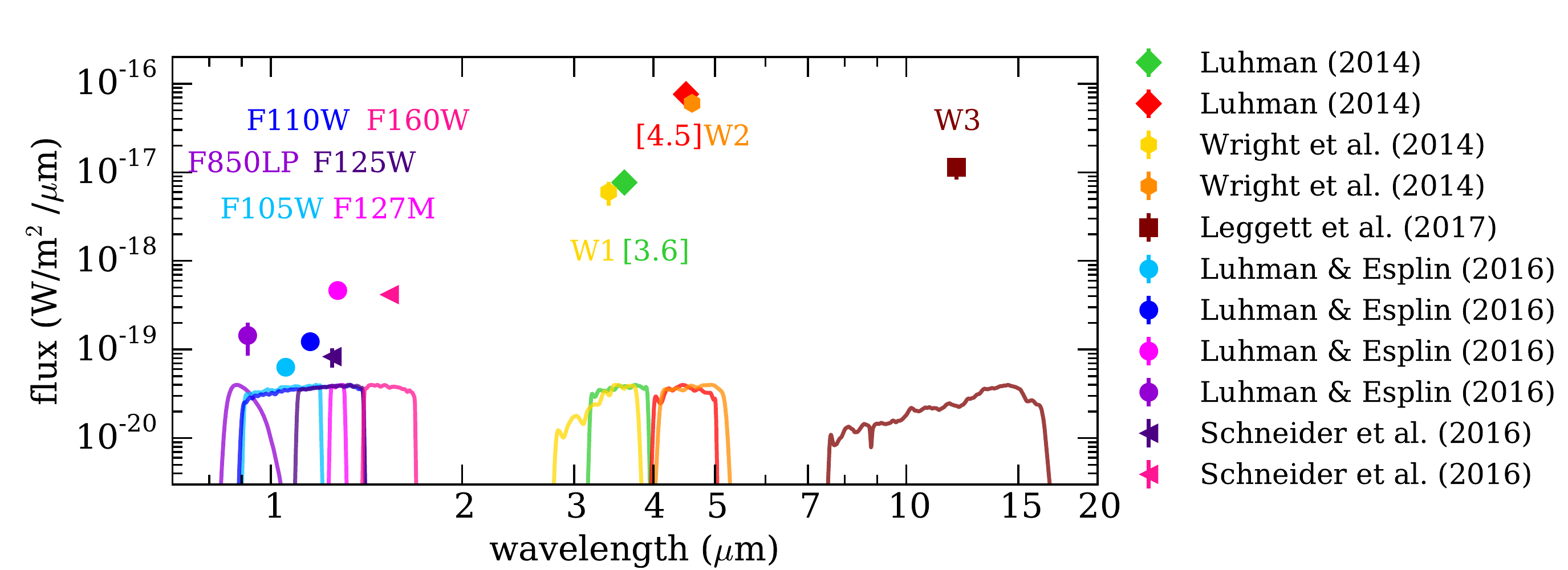}
 \caption{Space-based photometry of WISE 0855 from the literature, converted to flux density. Photometric measurements are shown as points; the corresponding filters are shown at the bottom, scaled for display on the plot. From blue to red, the filters are: \emph{HST}/WFC3 F850LP, F105W, F110W, F125W, F127M, F160W, \emph{WISE} W1 (yellow), \emph{Spitzer} IRAC1 (green), \emph{Spitzer} IRAC2 (red), \emph{WISE} W1 (orange), \emph{WISE} W3. }
\label{allphotom}
\end{figure*}

\citet{Schneider16} and \citet{Luhman16} aimed to further characterize WISE 0855's atmosphere using \emph{HST}/WFC3 and some additional ground-based observations. The object was detected in six optical and near-infrared filters on WFC3 and confirmed to be very faint at these wavelengths. Each study independently concluded that none of the available models matched the full SED of WISE 0855 simultaneously. The photometry is summarized in Table \ref{photometry_table} and Figure \ref{allphotom}. 

\begin{figure}[t]
\center \includegraphics[width=3.3in]{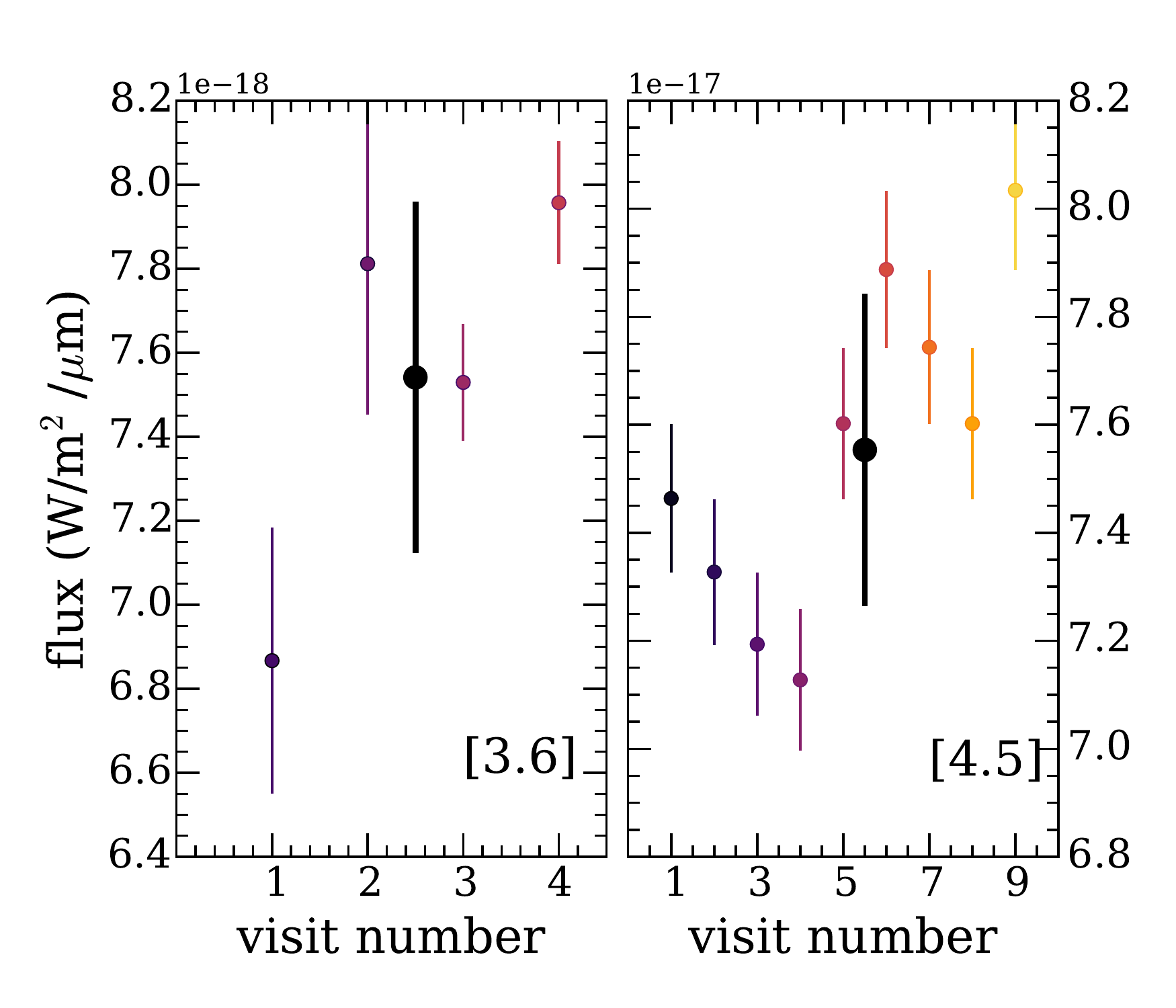}
 \caption{\emph{Spitzer} IRAC photometry of WISE 0855 \citep{Luhman16}, converted to flux density. Each smaller point indicates the measured photometry from a single visit; the center larger point shows the mean flux density with an error bar corresponding to the standard deviation. }
\label{spitzerphotom}
\end{figure}

Almost all of WISE 0855's emergent flux is at thermal infrared wavelengths longer than 2 \micron, as the Wien tail collapses at \teff<350 K \citep{Burrows03b}. In fact, while a 425 K Y dwarf is expected to emit about 10\% of its flux at wavelengths shorter than 2 \micron, for a $\sim$250 K Y dwarf like WISE 0855, only 0.1\% of its flux is emitted at these shorter wavelengths. WISE 0855 has been observed repeatedly using both \emph{WISE} and \emph{Spitzer} in the 3--15 \micron\ region. The thirteen \emph{Spitzer} photometry measurements are shown in Figure \ref{spitzerphotom}. 

These observations consistently show that, compared to a range of models, WISE 0855 is significantly brighter at 3--4 \micron\ than the models predict, by $\sim$1 mag \citep{Luhman14}. This discrepancy is not unique to WISE 0855; it is consistent across the population of late T and Y dwarfs cooler than 600 K \citep{Leggett13, Beichman14, Leggett17}. \citet{Leggett17} compared models with a variety of temperatures, gravities, clouds, disequilibrium carbon/nitrogen chemistry, and changes to the adiabatic slope \citep[see][]{Tremblin15}, but found that none of these physical properties explained the mid-infrared colors of late T and Y dwarfs. Models that adequately fit the mid-infrared photometry have not yet been developed by any groups. 

\citet{Esplin16} monitored WISE 0855 for photometric variability in \emph{Spitzer} IRAC1 and IRAC2 bands and detected peak-to-peak variability of 3--5\% in both bands within a 23-hour observation, at two different epochs. This variability indicates that the photosphere of WISE 0855 is heterogeneous in brightness, potentially caused by patchy clouds, inhomogeneous chemistry, or hot/cold spots. 

\begin{deluxetable}{lccr}

    \tablecaption{Photometry}
    \tablehead{\colhead{Filter} & \colhead{app. magnitude} & \colhead{$\lambda_{\rm center}$} & \colhead{reference} \\ 
\colhead{} & \colhead{(mag)} & \colhead{\micron} & \colhead{} } 

         \startdata
        $i^\prime$ & >27.2  & 0.780 & \citet{Luhman16} \\
        F850LP  & $26.85\substack{+0.31\\-0.44}$  & 0.832 & \citet{Luhman16} \\
        $z^\prime$ & >24.3  & 0.910 & \citet{Kopytova14} \\
        $Y$        & >24.4   & 1.021 & \citet{Beamin14} \\
        F105W  & $27.33\pm{0.19}$ & 1.0552 & \citet{Luhman16} \\
        F110W  & 26.71$\pm{0.19}$ & 1.1534 & \citet{Luhman16} \\
        F110W  & 26.471$\pm{0.13}$ & 1.1534 & \citet{Luhman16} \\
        F110W  & 26.00$\pm{0.12}$ & 1.1534& \citet{Luhman16} \\
        F125W  & 26.41$\pm{0.27}$ & 1.2486 & \citet{Schneider16} \\
        F127M  & 24.52$\pm{0.12}$ & 1.2740 & \citet{Luhman16} \\
        F127M  & 24.49$\pm{0.11}$ & 1.2740& \citet{Luhman16} \\
        F127M  & 24.36$\pm{0.09}$ & 1.2740 & \citet{Luhman16} \\
        $J3$        & $24.8\substack{+0.53 \\ -0.35}$ & 1.3 & \citet{Faherty14b} \\
        F160W  & 23.86$\pm{0.03}$ & 1.5369 & \citet{Schneider16} \\
        CH$_4$cont & $23.2\pm{0.2}$  & 1.575 & \citet{Luhman16} \\ 
        $H$        & >22.7   & 1.633 & \citet{Wright14} \\
        $K_s$      & >18.6   & 2.146 & \citet{McMahon13}  \\
        $W$1       & 17.82$\pm{0.33}$  & 3.3526 & \citet{Wright14}  \\
        IRAC1      & 17.44$\pm{0.05}$   & 3.550 & \citet{Luhman16} \\ 
        IRAC1      & 17.30$\pm{0.05}$   & 3.550 & \citet{Luhman16} \\ 
        IRAC1      & 17.34$\pm{0.02}$   & 3.550 & \citet{Esplin16} \\ 
        IRAC1      & 17.28$\pm{0.02}$   & 3.550 & \citet{Esplin16} \\ 
        IRAC2      & 13.88$\pm{0.02}$   & 4.493 & \citet{Luhman16} \\
        IRAC2      & 13.90$\pm{0.02}$   & 4.493& \citet{Luhman16} \\ 
        IRAC2      & 13.92$\pm{0.02}$   & 4.493& \citet{Luhman16} \\ 
        IRAC2      & 13.93$\pm{0.02}$   & 4.493& \citet{Luhman16} \\ 
        IRAC2      & 13.86$\pm{0.02}$   & 4.493& \citet{Luhman16} \\ 
        IRAC2      & 13.82$\pm{0.02}$   & 4.493& \citet{Luhman16} \\ 
        IRAC2      & 13.84$\pm{0.02}$   & 4.493& \citet{Esplin16} \\ 
        IRAC2      & 13.86$\pm{0.02}$   &4.493 & \citet{Luhman16} \\ 
        IRAC2      & 13.80$\pm{0.02}$   &4.493 & \citet{Esplin16} \\
        $W$2       & 14.02$\pm{0.05}$  & 4.6028 & \citet{Wright14}  \\
        W3      & 11.9$\pm{0.3}$   & 11.5608 & \citet{Leggett17} \\
\enddata
    \label{photometry_table}
\end{deluxetable}

\clearpage 

\subsection{Previous Spectroscopic Observations of WISE 0855} \label{prev_specs}

\citet{Skemer16} observed the first spectrum of WISE 0855 using the GNIRS instrument on the Gemini North telescope in M band, from 4.5--5.1 \micron. The spectrum is dominated by water vapor features; it shows muted amplitude features and flat continuum shape consistent with a cloud deck in the atmosphere. The spectrum also revealed an unexpected lack of phosphine (PH$_3$) in contrast to Jupiter's atmosphere which has a strong PH$_3$ feature at these wavelengths. 

\subsection{This Work}

Here, we present a spectrum of WISE 0855 in the L band, from 3.4 to 4.14 \micron. This wavelength region is complementary to the M band since it probes the long wavelength portion of a strong CH$_4$ absorption feature (and therefore carbon abundance), while the M band probes H$_2$O features. This region is also sensitive to water clouds and PH$_3$ absorption. We combine this L band measurement with the M band spectrum from \citet{Skemer16} to provide insight into the mid-infrared spectrum of WISE 0855. We generate a new set of cold atmospheric models, with a variety of compositions and including the effect of clouds, in order to fit the observed spectra and photometry of WISE 0855.

%\clearpage
\section{Methods}

\subsection{Observations}

We obtained an $L$-band spectrum of \scitarg\ using the GNIRS spectrograph \citep{Elias06} on the Gemini-North
Telescope (program GN-2017A-FT-6). Gemini North is operated in queue mode, which allowed us to observe WISE 0855 during clear, dry, calm conditions over the course of several nights. We used the 10.44~lines~mm$^{-1}$ grating with the 0\farcs05 pix$^{-1}$ camera and a 0\farcs675 x 49$''$ slit, resulting in a 3.2--4.14~$\mu$m spectrum. The spectral resolving power varies linearly with wavelength from $\sim$240 at 3.2 \micron\ to $\sim$310 at 4.1 \micron\ with an average of $R~\approx$~275. We used the "Bright Object" readout mode for GNIRS which includes one low noise read and 16 digital averages, resulting in a minimum integration time of 0.55~s. To balance observing efficiency with saturation at long wavelengths, we chose an integration time of 5~s. For observations of WISE 0855, each exposure was 5~s~$\times$~11 coadds. For observations of telluric standards, each exposure was 5~s~$\times$~4 coadds. All observations were taken using a 6\arcsec\ ABBA nod pattern to enable subtraction of the telescope and sky background.  Nine ABBA patterns (36 total exposures, for 1980~s of total integration time) composed each set of WISE 0855 observations. A single ABBA pattern (4 total exposures, for 80~s of total integration time) was obtained for each telluric standard.

Our queue observations (summarized in Table \ref{observingtable}) were conducted during 4 nights in March 2017. Each night, our observations started with the observation of a B8V type telluric standard, HIP 39898, followed by one $\sim$45~minute set of WISE 0855 observations. WISE 0855 was then reacquired, and another set of WISE 0855 observations were obtained, followed by observation of HIP~49900, an A0V telluric standard. The total integration time on WISE 0855 was 4.4 hours over 4 nights. All observations of WISE 0855 were obtained within 1.5 hours of meridian transit. Thus, the airmass at the time of our observations was optimal, and the airmass match between WISE 0855 and our telluric standards was excellent, with only a slight ($\le$0.04) difference in airmass between WISE 0855 and associated telluric standard observations. 

WISE 0855 has a high proper motion of 8.118$''$ per year \citep{Luhman16}. The location of WISE 0855 at the time of the observations was calculated by propagating its proper motion and parallax \citep{Luhman16}, which provides a location accurate to better than $\sim0.1''$, which is the blind-offset pointing precision of Gemini. As in \citet{Skemer16}, we used a blind acquisition technique to place WISE 0855 in the slit, using an offset from a nearby star on the sky with a calibrated astrometric location \citep{Faherty14b}. The chosen 0\farcs675 slit assures us that the observation is not subject to significant slit losses.

Observing conditions for the Gemini queue were restricted to 0.50$''$ seeing or better in L band (70th percentile on Mauna Kea), cloudless (50th percentile), and 3mm or less precipitable water vapor (80th percentile).

\begin{deluxetable}{lcrrlc}
\tablecaption{Observations\label{observingtable}}
%\tablewidth{0pt}
%\tablecolumns{2}
\tablehead{
  \colhead{Date} &
  \colhead{{\it WISE}~0855} &
  \colhead{IQ} &
  \colhead{WV} &
  \colhead{Telluric} &
  \colhead{Telluric} \\
  \colhead{(UT)} &
  \colhead{Airmass} &
  \colhead{} &
  \colhead{} &
  \colhead{} &
  \colhead{Airmass}
}
\startdata
2017 Mar 04 & 1.17 & 70\% & \nodata  & HIP~39898 & 1.20 \\
2017 Mar 04 & 1.13 & 70\% & \nodata  & HIP~49900 & 1.15 \\
2017 Mar 27 & 1.16 & 20\% & 50\% & HIP~39898 & 1.19 \\
2017 Mar 27 & 1.13 & 20\% & 20\% & HIP~49900 & 1.15 \\
2017 Mar 28 & 1.14 & 70\% & 20\% & HIP~39898 & 1.18 \\
2017 Mar 28 & 1.13 & 70\% & 50\% & HIP~49900 & 1.13 \\
2017 Mar 29 & 1.16 & 70\% & 50\% & HIP~39898 & 1.19 \\
2017 Mar 29 & 1.13 & 70\% & 50\% & HIP~49900 & 1.14 \\
\enddata
\end{deluxetable}

\subsection{Data Reduction}

We first rotate the raw exposures by 90 degrees, so that the spatial direction is roughly aligned with image columns and the dispersion direction is roughly aligned with rows. We then flag any pixels with values greater than 10,000 per coadd as nonlinear. Rather than correct for non-linearity, we carry these flags throughout the reduction process. We scale each image by its median and then create nod-subtracted (A-B) pairs from adjacent A and B images.

To spatially and spectrally rectify our data, we use a modified version of the REDSPEC package\footnote{\url{http://www2.keck.hawaii.edu/inst/nirspec/redspec}}. We create a spatial map for each telluric standard using an A-B pair and REDSPEC's {\tt spatmap} procedure. We use this spatial map to remap each exposure so that the dispersion direction lies along detector rows. To wavelength calibrate and spectrally rectify our exposures, we use sky emission present in our telluric standard exposures. The brightness of the thermal background makes fitting individual sky lines difficult. Thus, we compare the observed sky emission to a model of the sky background for Mauna Kea provided on the Gemini Observatory website\footnote{\url{http://www.gemini.edu/sciops/telescopes-and-sites/observing-condition-constraints/ir-background-spectra}} for an airmass of 1.0 and precipitable water vapor content of 1.0~mm. We first smooth the sky model to R$\sim$273, and then run both the sky model and our observed sky emission through a high-pass filter to remove thermal emission. We create a 2nd-order polynomial wavelength solution for our observed sky spectrum using barycentric Lagrangian interpolation \citep{Berrut04,Waring79}. We use IDL's {\tt AMOEBA} to solve for the wavelengths of the first, middle, and last pixels of our observed sky spectrum which minimizes residuals between our model and observed sky spectra. To create a spectral map, we use {\tt AMOEBA} to solve for the dispersion solution at 5 different rows (i.e.~different spatial positions), which we then use to remap each exposure so that each image column corresponds to a single wavelength. The spectral and spatial maps determined for each set of telluric standard data are likewise applied to adjacent observations of \scitarg.

For observations of \scitarg\, we remove residual sky lines by subtracting the median from each column in the rectified, nod-subtracted A-B pairs. We then combine the rectified, nod-subtracted A-B pairs with 3-$\sigma$ clipped mean at each pixel. The uncertainty of each pixel in the stacked image is the standard deviation of the mean. Our uncertainties agree with expectations from Poisson statistics.

We extract a spectrum for each nod position in our stacked A-B images. We first determine the row corresponding to each nod position by taking the median along each row for columns corresponding to wavelengths of 3.70--4.14~$\mu$m (where flux from the object is significant). We smooth this spatial cut by 7 pixels (0\farcs35), and then fit two Gaussians to the smoothed spatial cut. For each nod, we extract the spectrum using an aperture centered at the peak of the Gaussian and having an aperture radius of 6--9.5 pixels (0\farcs3--0\farcs475), which typically match the seeing and visually encompassed the bulk of the object's flux in the spatial cut. We subtract any sky background residuals by fitting a line to background regions 75 pixels in width and starting 3 aperture radii away from the edge of the aperture. If a pixel in the aperture is flagged as non-linear in more than 10\% of the raw images, the extracted spectrum at that wavelength is also flagged. For each night of observation, our reduction results in an A and B spectrum for each observation set, or four extracted spectra of \scitarg. For our telluric standards, we use REDSPEC to extract the final spectra, tracking uncertainties using Poisson statistics.

\begin{figure}[t]
\center \includegraphics[width=3.5in]{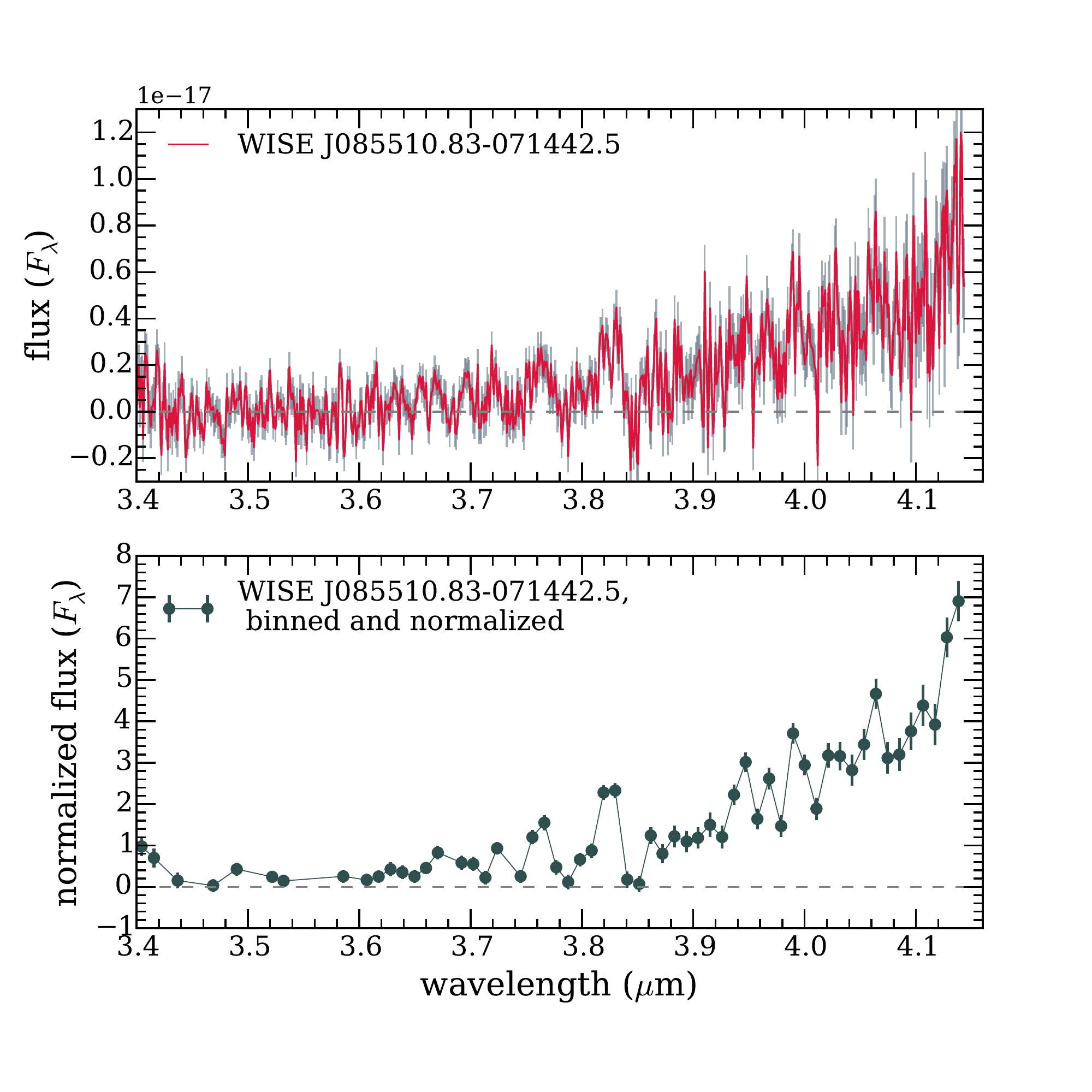}
 \caption{Gemini/GNIRS spectrum in L band of WISE 0855. The spectrum increases in brightness with wavelength and shows a number of absorption features. The top panel shows the unbinned spectrum (red) with error bars (gray). The bottom panel shows the binned spectrum, which we normalize (from 3.4--4.12 \micron) because of the flux calibration uncertainty associated with blind-offset slit misalignment. }
\label{alone}
\end{figure}

\begin{figure*}[t]
\vspace{-0.05in}
\center \includegraphics[width=4.7in]{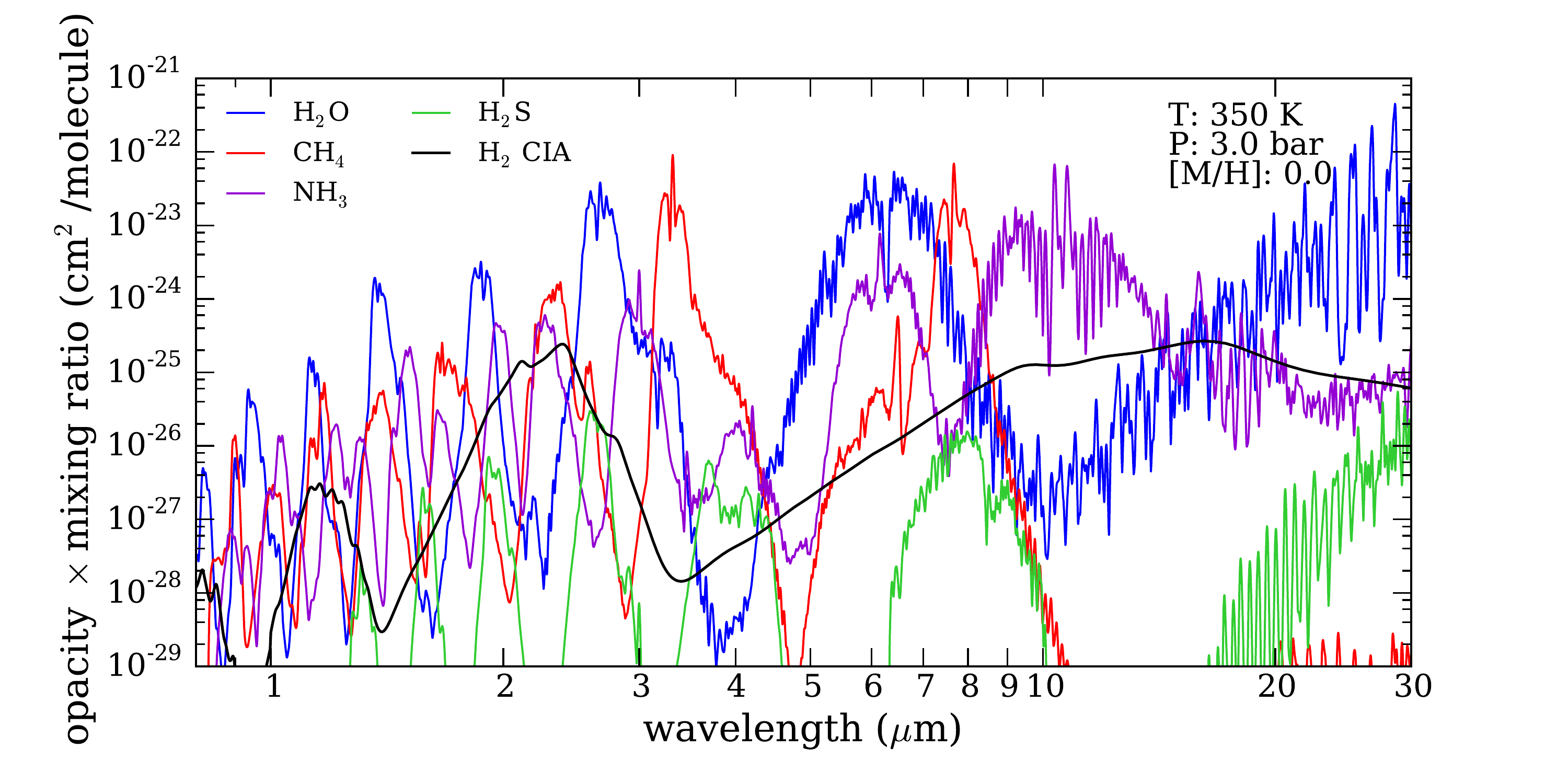}
%\vspace{-0.1in}
 \caption{Cross sections of a number of abundant molecules. The cross sections are scaled by the mixing ratio in chemical equilibrium at a temperature of 350 K and pressure of 3 bar, which is approximately the temperature of the mid-infrared photosphere of a 250 K object. The dominant absorbers across the spectrum are H$_2$O, CH$_4$, and NH$_3$. }
\label{opacities}
\end{figure*}

\begin{figure}[h]
\center \includegraphics[width=3.6in]{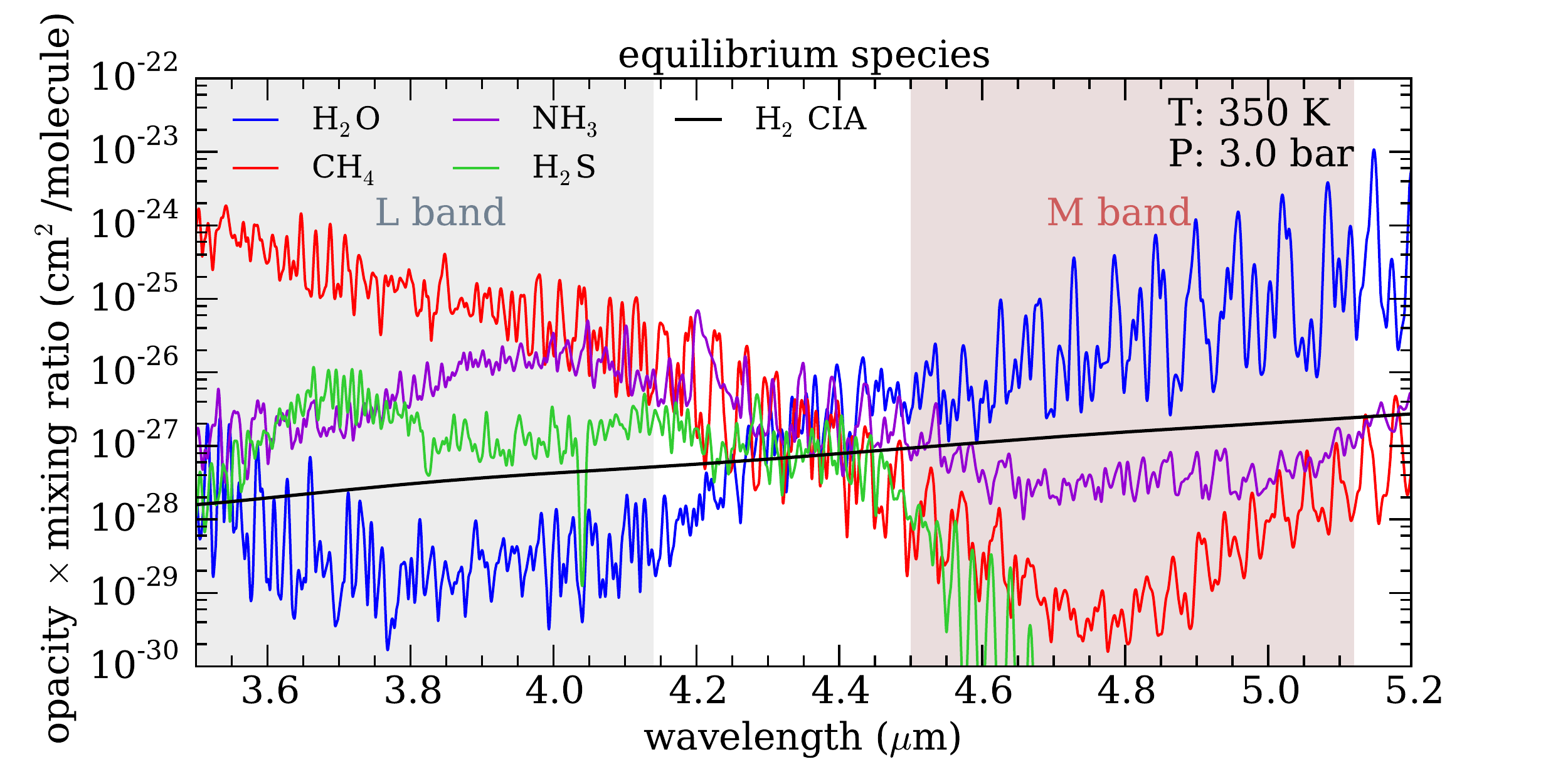}
\vspace{-0.4in}
\center \includegraphics[width=3.6in]{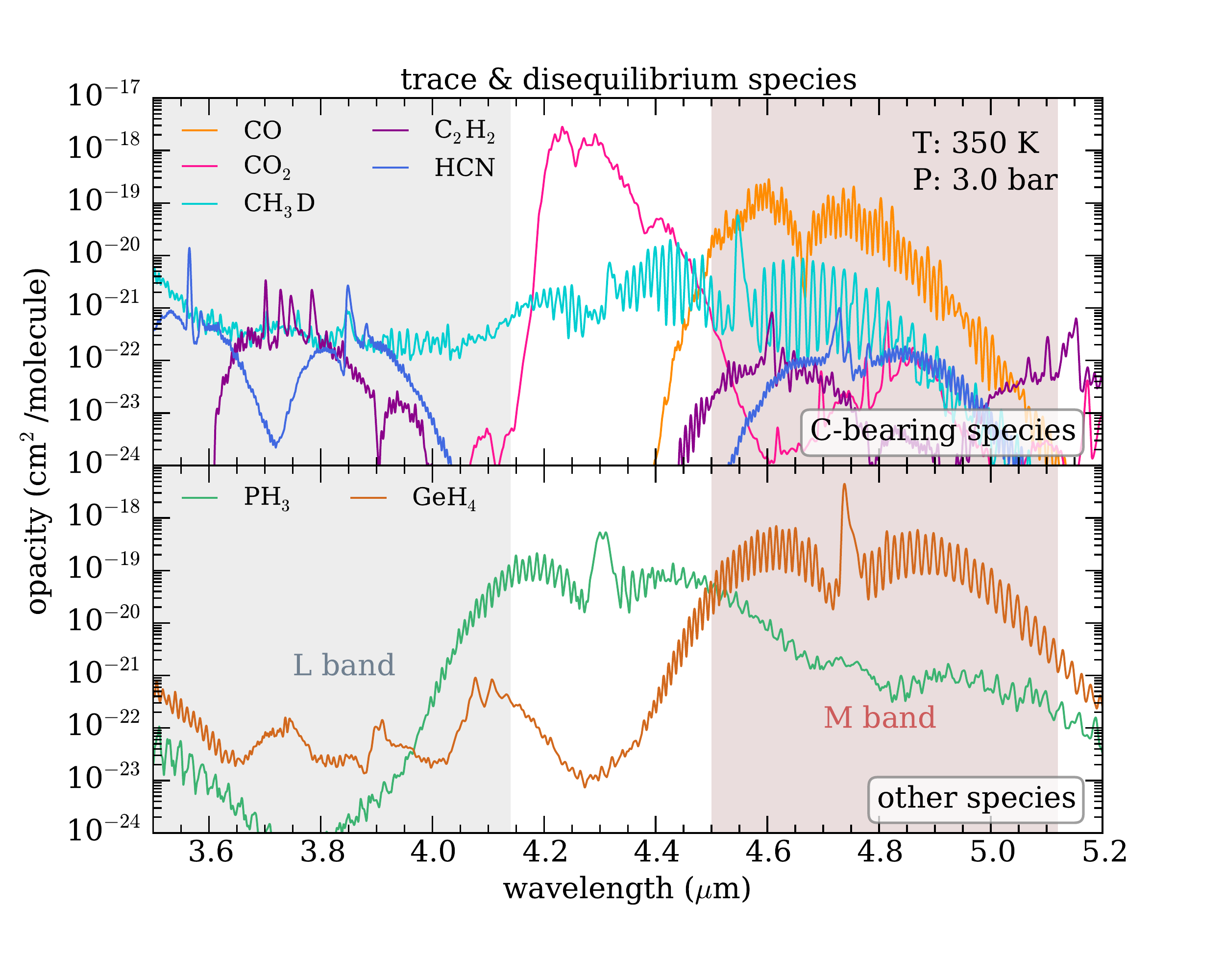}
 \caption{Cross sections of molecular species in the mid-infrared. The top panel shows abundant species in chemical equilibrium, scaled by their mixing ratios assuming equilibrium (at T=350 K, P=3 bar, [M/H]=0.0, the approximate temperature and pressure of the 3--5 \micron\ photosphere of a 250 K brown dwarf). L band is dominated by methane and ammonia, while M band is dominated by water. The bottom panel shows potential trace species (those \emph{not} abundant in equilibrium at this temperature and pressure), not scaled by abundance. In both panels, the wavelengths probed in L and M bands are shaded. }
\label{opacities_zoom}
\end{figure}

We correct each extracted spectrum of \scitarg\ for telluric absorption by creating a telluric spectrum from an adjacent telluric standard. To create a telluric absorption spectrum, we first remove the Pfund $\gamma$ and Brackett $\alpha$ lines in the telluric standard spectrum by fitting and subtracting Gaussian fits to the lines. We then divide the telluric standard spectrum by an appropriate blackbody for the temperature and $V$-band magnitude of the telluric star. In total, we had 16 telluric-corrected spectra of \scitarg, which we then combine using robust weighted mean \citep{Cushing04}.

The spectrum is shown in Figure \ref{alone}. We bin the final spectrum in wavelength using a weighted average to a spectral resolution of $\sim$250 ($\sim$10 pixels per bin). The final binned spectrum, shown in the bottom panel of Figure \ref{alone}, has a mean S/N per bin of $\sim$6 from 3.53 to 4.14\micron.

\begin{figure*}[thb]
\center \includegraphics[width=6.0in]{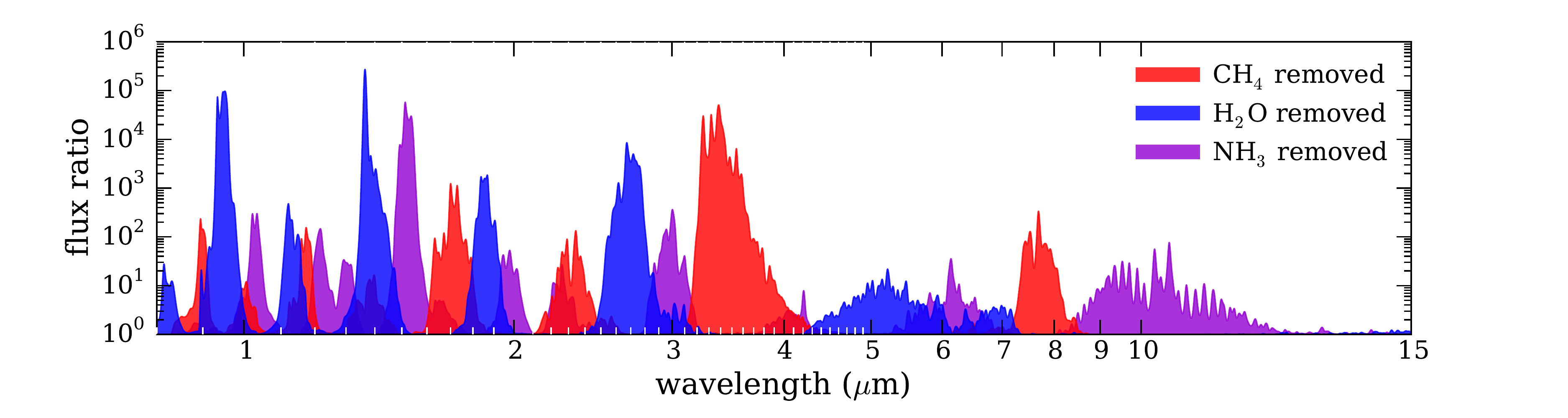}
 \caption{The sensitivity of the spectrum to each wavelength. For a model with T$_{\rm eff}$=250 K, log g=4.0, [M/H]=0.0, and C/O=0.6, water, methane, and ammonia are removed individually, keeping the P--T profile the same; the spectrum with each molecule removed is divided by the standard spectrum. A 250 K object's spectrum is dominated by these three species.  }
\label{elements}
\end{figure*}

\subsection{Model Atmospheres}

We generate new atmospheric models to compare to the measured photometry and spectra of WISE 0855. We run models with a small range in effective temperatures of 250--270 K and log g=4.0. WISE 0855's effective temperature has been constrained to a small range in previous works using models with a wider range of temperatures \citep[e.g.,][find that the best fit models are in this \teff\ range]{Luhman14}. We find comparing to previous larger model grids at solar metallicity \citep[e.g.,][]{Morley14a} that models around 250--270 K provide the best fits, and therefore focus on this range for our detailed model grid presented here. 

We include models at a variety of metallicities, focusing on those around solar metallicity ([M/H]=0.0 and [M/H]=$-0.25$), since we expect most objects in the solar neighborhood to have solar-like metallicities. We also include three different C/O ratios: 0.6 (solar), 0.3, and 0.15. To model different C/O ratios, we keep the oxygen abundance constant and vary the carbon abundance. 

Solar metallicity ([M/H]=0.0) corresponds to a deep H$_2$O mixing ratio of 8.2$\times10^{-4}$ and CH$_4$ mixing ratio of 4.7$\times10^{-4}$. [M/H]=$-$0.25 corresponds to a deep H$_2$O mixing ratio of 4.6$\times10^{-4}$ and a CH$_4$ mixing ratio of 2.6$\times10^{-4}$. Other carbon and oxygen-bearing species like CO and CO$_2$ are also included; neither are abundant (mixing ratios smaller than 10$^{-20}$) in the photosphere (1 bar), but can be present in the deep atmosphere. For example, at solar metallicity the CO and CO$_2$ mixing ratios at $\sim$300 bar are 4.6$\times10^{-4}$ and 5$\times10^{-7}$ respectively.

To generate model atmospheres, we calculate the temperature structures assuming radiative-convective equilibrium. These models are extensively described in \citet{McKay89, Marley96,Marley99,Marley02, Burrows97, Fortney08a, Saumon08, Morley12, Morley14a}. Our opacity database for gases is described in \citet{Freedman08, Freedman14}. We include updates to the opacities for a number of gas species. The abundances of molecular, atomic, and ionic species are calculated using a modified version of the NASA CEA Gibbs minimization code \citep{[see][] McBride92}. Further details on the opacities and chemical equilibrium are described in Marley et al. (in prep.).

For these cold models, we include water ice, sulfide (Na$_2$S, ZnS, MnS), and salt (KCl) clouds. We use a modified version of the \citet{AM01} cloud model to include these species, as described in \citet{Morley12, Morley13, Morley14a}. Cloud material in excess of the saturation vapor pressure of the limiting gas is assumed to condense into spherical, homogeneous cloud particles. Cloud particle sizes and vertical distributions are calculated by balancing transport by advection with particle settling. The free parameter \fsed\ represents the sedimentation efficiency of cloud material (see \citet{AM01} for details). Lower \fsed\ values lead to more vertically extended clouds of smaller particles. We calculate the effect of cloud opacity using Mie theory, assuming spherical particles. Optical properties of sulfide, salt, and water ice clouds are from a variety of sources and presented in \citet{Morley12, Morley14a}. 

We use a radiative transfer model developed in \citet{Morley15} (see Appendix of that work) to calculate the thermal emission at moderate spectral resolution. Briefly, this model includes the C version of the open-source radiative transfer code \texttt{disort} \citep{Stamnes88, Buras11}, which uses the discrete-ordinate method to calculate intensities and fluxes in multiple-scattering and emitting layered media. We also use this radiative transfer model to calculate the effect of changing the abundance of trace species without changing the temperature structure of the atmosphere (e.g., in Section \ref{ph3}).

\subsubsection{Partial vs. Homogeneous Clouds}

One substantial difference from the approach taken in \citet{Morley14a} is that we consider homogeneous clouds rather than `partial' (50\%) cloud cover, converging pressure--temperature profiles in radiative-equilibrium with a fully cloudy atmosphere. Homogeneous clouds have a stronger effect on the mid-infrared spectrum. 

In \citet{Morley14a}, the majority of models were 50\% cloudy, 50\% clear, which ensured that pressure--temperature profiles could converge for all objects in that temperature range. The most numerically unstable region---in terms of converging the temperature profile---is from $\sim$300 to  $\sim$375 K, right as the water ice clouds are beginning to form. This occurs because the clouds form high in the atmosphere, and the temperature profile ends up very close to, and parallel to, the condensation curve. Small changes in temperature therefore give a large change in cloud opacity and gas opacity. For colder objects, the water clouds form deeper, in a region where the temperature profile is steeper and no longer parallel to the condensation curve. The same small deviations in temperature have a smaller impact on gas and cloud opacity, and the models converge somewhat more easily. We run homogeneous models for this new work, more similar to all of the previous cloudy brown dwarf models from our group (\citet{Marley02} and \citet{Saumon08} L dwarf models, \citet{Morley12} T dwarf models), with the patchy clouds from \citet{Marley10} and \citet{Morley14a} being the exceptions to the rule in modeling approach.

\subsubsection{Molecular Opacities in Y Dwarf Atmospheres}

Cold (\teff<300 K) substellar atmospheres are dominated by opacity from molecular species, including CH$_4$, H$_2$O, and NH$_3$ \citep{Burrows03b}. Cross sections for a variety of molecules are shown in Figures \ref{opacities} and \ref{opacities_zoom} for the approximate pressure and temperature for the mid-infrared photosphere of a $\sim$250 K Y dwarf like WISE 0855. In Figure \ref{opacities}, the cross section across the near- and mid-infrared are shown for species in chemical equilibrium; in Figure \ref{opacities_zoom}, the cross sections are shown for the 3.5--5.2 \micron\ region probed by the observed spectra. The dominant source of opacity in M band is water vapor, while the L band is dominated by methane and, at the reddest wavelengths, ammonia.

 We also consider species that are not present at high abundances in chemical equilibrium, but which could be present in these atmospheres. These include species that might increase in abundance due to disequilibrium processes such as vertical mixing and photochemistry as well as species predicted to be present in equilibrium in trace amounts. For example, if Jupiter's atmosphere were in chemical equilibrium, phosphorous would be in the form of P$_4$O$_6$ \citep{Fegley94, Visscher06}. Instead, Jupiter's atmosphere is abundant in PH$_3$ with a mixing ratio around 0.8 ppm (about 3$\times$ solar) \citep{Irwin98}. This indicates that vertical mixing from deep, hot layers of Jupiter's atmosphere must occur on a faster timescale than PH$_3$ can be converted into P$_4$O$_6$, allowing us to probe mixing in Jupiter's atmosphere. Other trace species can be used to measure atmospheric dynamics in a similar way.

In Figure \ref{opacities_zoom}, we include the carbon-bearing species CO, CO$_2$, HCN, CH$_3$D and C$_2$H$_2$. We also include phosphine (PH$_3$), which we will discuss in more detail in Section \ref{ph3}, and germane (GeH$_4$), both of which are detected in Jupiter's atmosphere in the mid-infrared.

CO is a strong absorber in M band; CO$_2$ absorbs strongly between L and M band, largely preventing ground-based observations of faint objects from 4.15--4.5 \micron\ from within Earth's CO$_2$-rich atmosphere; PH$_3$ has a strong absorption feature centered at 4.3 \micron, between L and M band, and GeH$_4$ has a strong feature at 4.75 \micron. HCN and C$_2$H$_2$ have lower amplitude features across the mid-infrared. CH$_3$D may be accessible with \emph{JWST} at 4.55 \micron\ \citep[Morley et al., in prep.]{Skemer16} and would provide constraints on WISE 0855's mass and formation processes.

The sensitivity of the model thermal emission spectra to the three dominant molecular opacity sources (CH$_4$, H$_2$O, NH$_3$) is shown in Figure \ref{elements}. We show the ratio between a model spectrum calculated assuming chemical equilibrium and one in which the opacity of each molecule is removed, keeping the P--T profile constant. This provides an illustration of which wavelengths are controlled by each species.

\section{Results}

\subsection{WISE 0855's L band Spectrum}
%\section{WISE 0855's L band Spectrum}

\begin{figure}[t]
\center \includegraphics[width=3.5in]{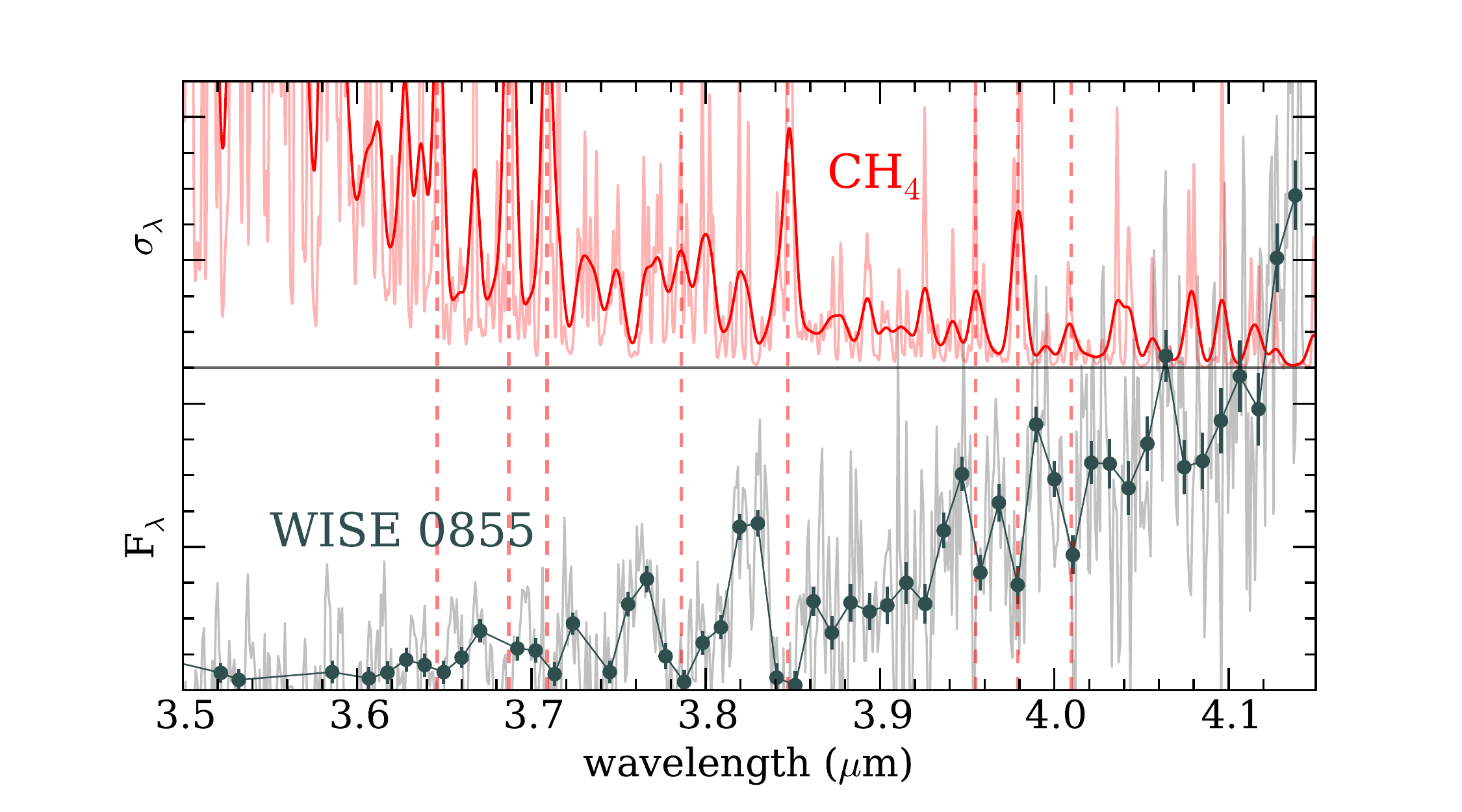}
 \caption{The normalized spectrum of WISE 0855 compared to cross sections of CH$_4$ from 3.5 to 4.14 \micron. The methane cross sections are shown at higher resolution (R$\sim$1500, light red) and lower resolution (R$\sim$300, solid red). Vertical dashed red lines centered on methane absorption bands are shown to guide the eye. All major absorption features seen in WISE 0855's L band spectrum correspond in wavelength with molecular bands of CH$_4$. We conclude the L band spectrum shows strong evidence of methane absorption. }
\label{allch4}
\end{figure}

\begin{figure}[th]
\center \includegraphics[width=3.7in]{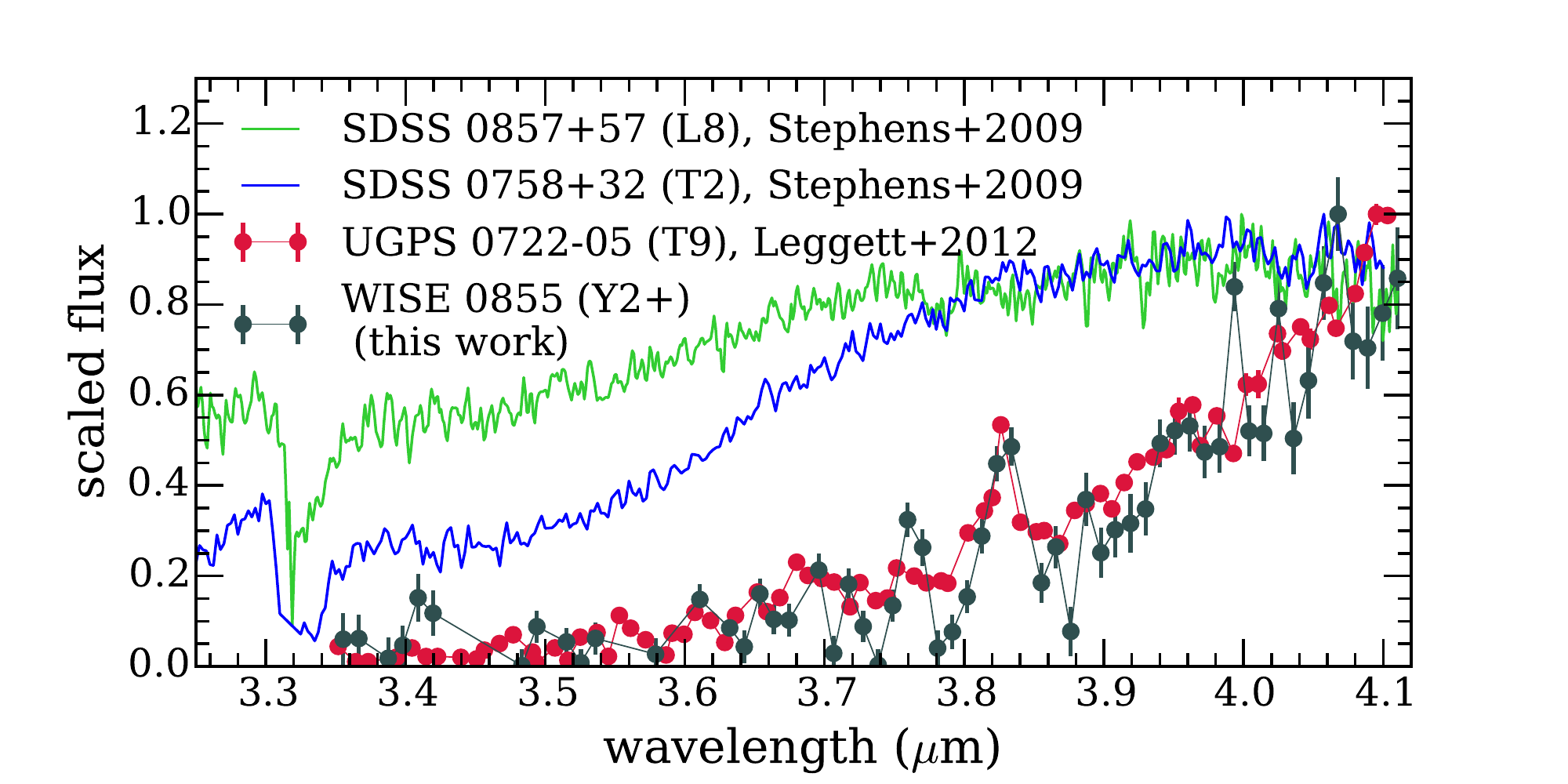}
\vspace{-0.4in}
\center \includegraphics[width=3.7in]{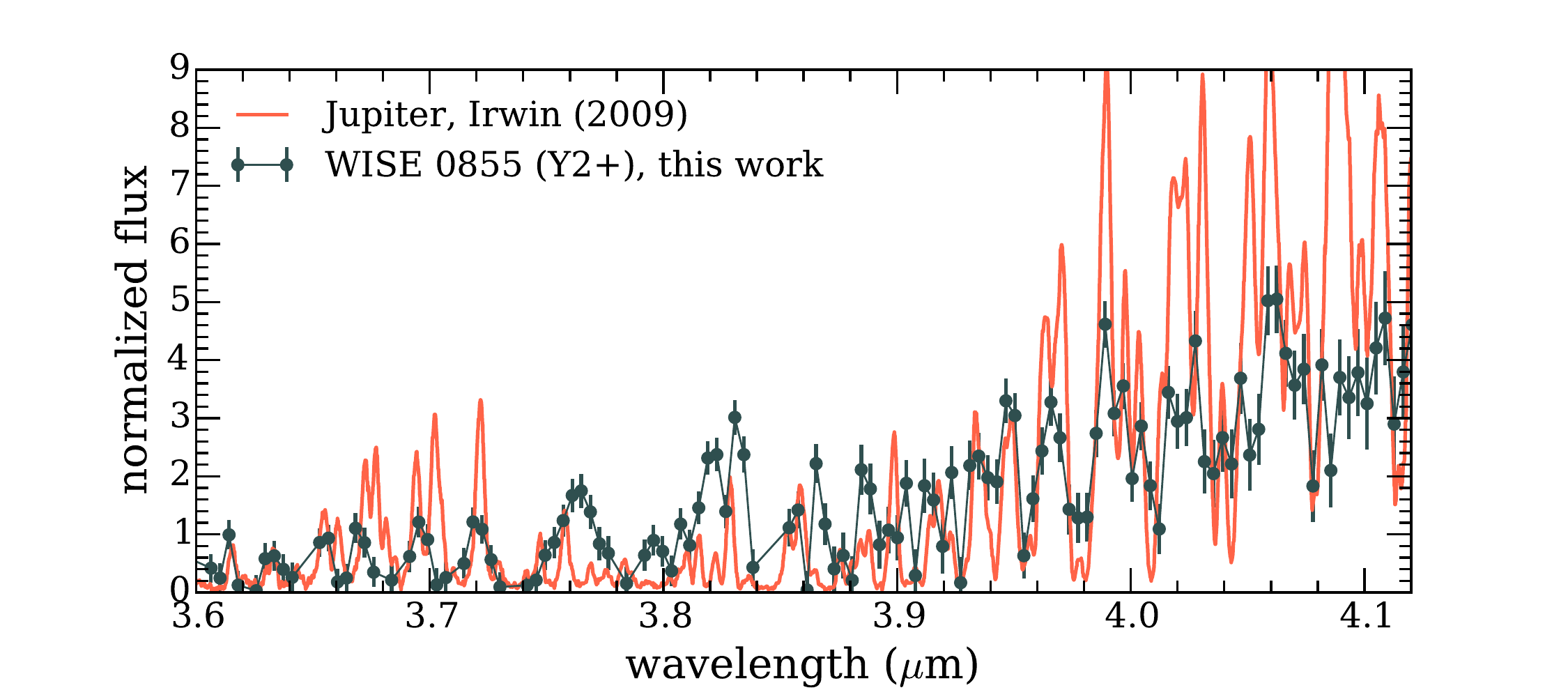}
\vspace{-0.4in}
\center \includegraphics[width=3.7in]{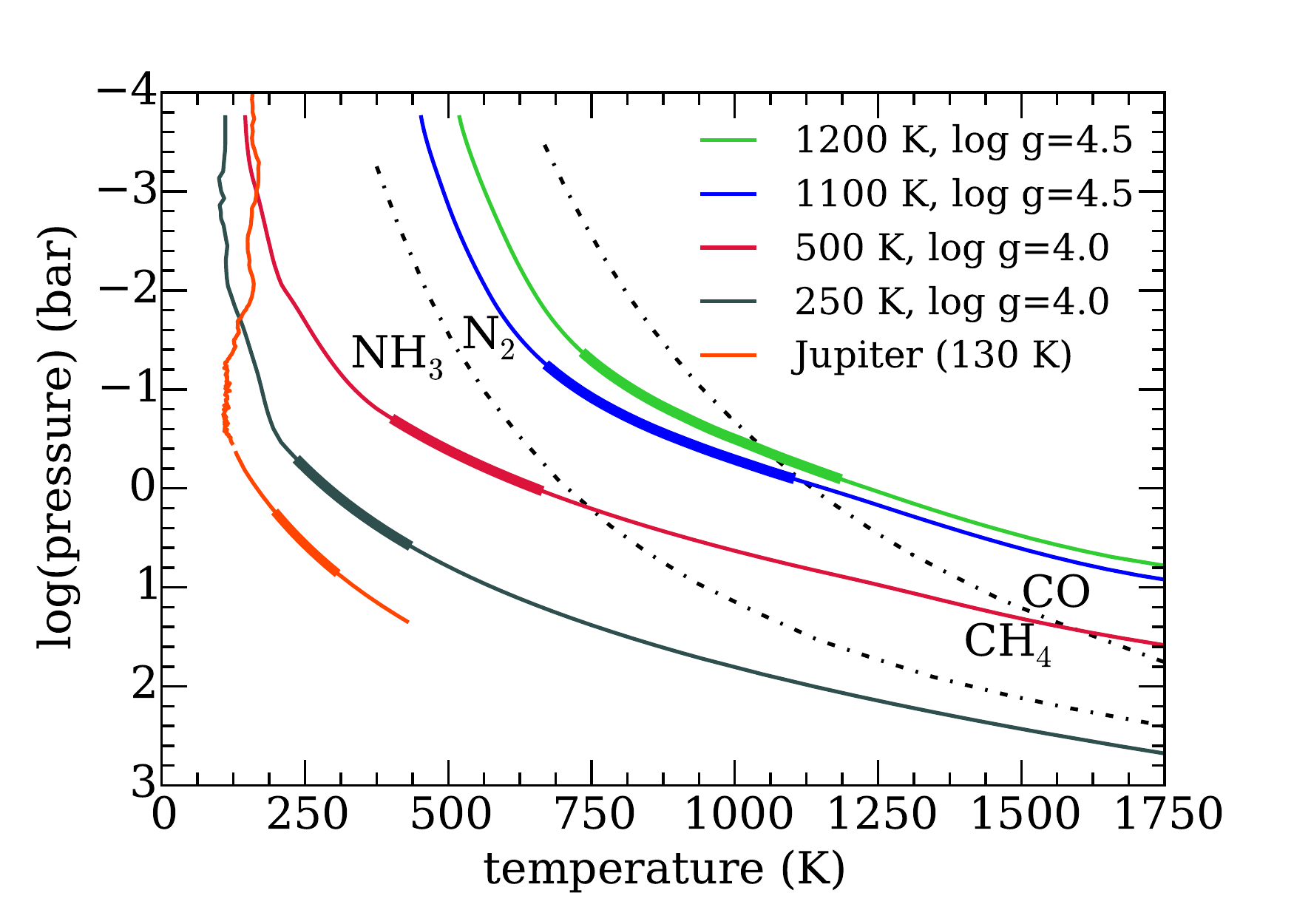}
 \caption{The L band spectrum of WISE 0855 compared to other brown dwarfs and Jupiter. The top panel shows four spectra of brown dwarfs with spectral types L8, T2, T9, and Y2+ \citep{Knapp04, Lucas10}. We scale the flux to have a maximum value of 1 between 3.3 and 4.1 \micron. Later spectral types show a more pronounced slope from methane absorption, and WISE 0855 has the highest amplitude features. The middle panel shows WISE 0855's spectrum (binned to a slightly higher spectral resolution than in the top panel) compared to Jupiter's spectrum, which is dominated by reflected light at these wavelengths. The bottom panel shows model pressure-temperature profiles at four temperatures that correspond approximately to the spectral types in the top panel as well as Jupiter's observed temperature profile \citep{Seiff98}. The approximate location of the 3.4--4.1 \micron\ photosphere is shown as a thick line. The two dash-dot lines show the temperatures and pressures where NH$_3$ and N$_2$, and CH$_4$ and CO have equal abundances. }
\label{sequence}
\end{figure}

\begin{figure*}[t]
\center \includegraphics[width=5.5in]{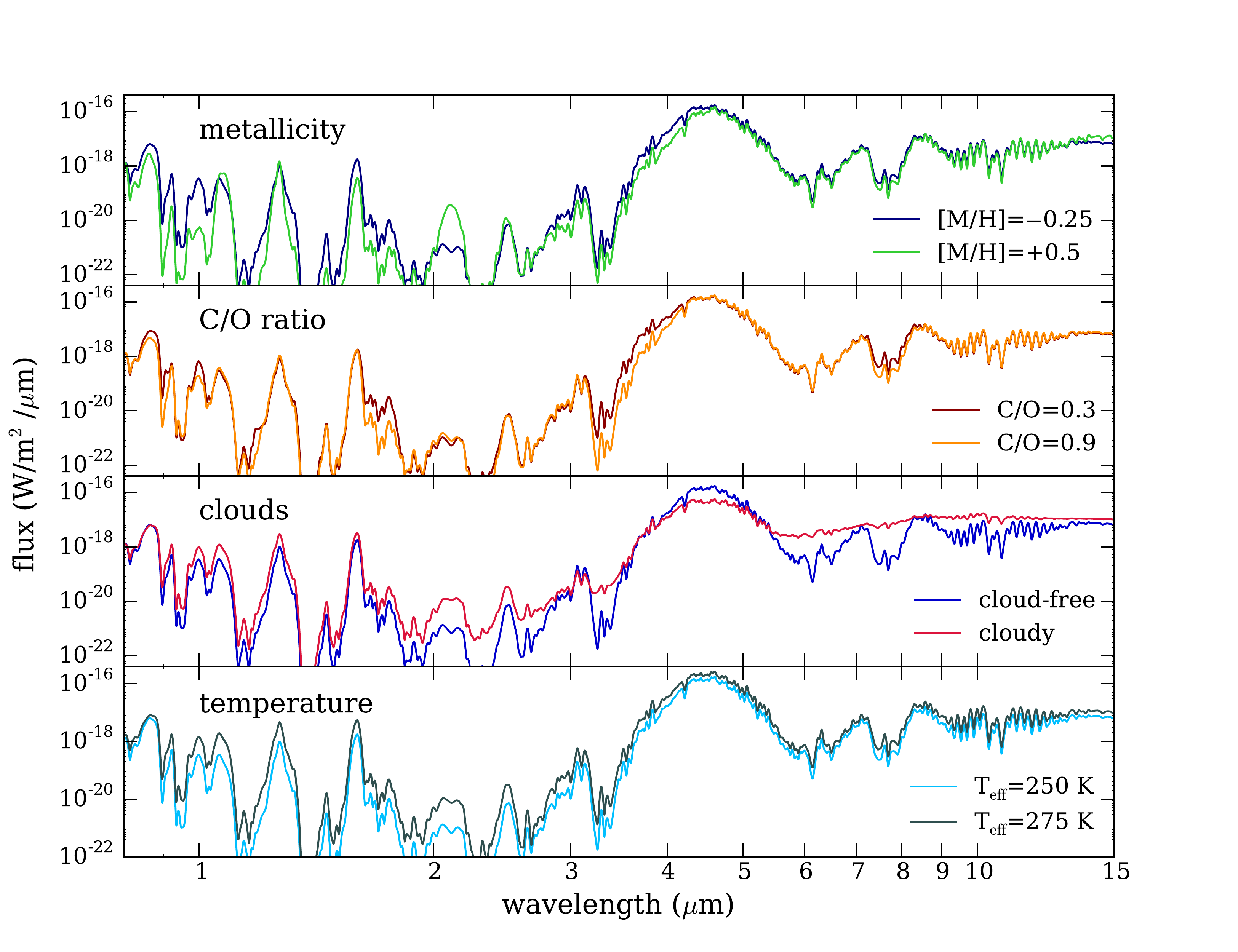}
 \caption{Model spectra illustrating the effect of changing four different model parameters. From top to bottom, these panels show models with different metallicity, C/O ratio, clouds, and temperature. All models have surface gravity of 100 m/s$^2$. Top panel: 250 K, cloud-free, C/O ratio of 0.6, and [M/H] of -0.25 and +0.5. 2nd panel: 250 K, cloud-free, [M/H] of -0.25, and C/O ratio of 0.3 and 0.9. 3rd panel: 250 K, C/O ratio of 0.6, [M/H] of -0.25, and cloud-free and cloud parameter $f_{\rm sed}=4$. Bottom panel: cloud-free, [M/H] of -0.25, C/O ratio of 0.6, and temperature of 250 K and 275 K. }
\label{model_spectra}
\end{figure*}

The L band spectrum of WISE 0855 shown in Figure \ref{alone} increases in brightness from 3.5 to 4.14 \micron\ as the cross section of methane decreases (see Figure \ref{opacities_zoom}). It also shows a number of absorption features, in particular at $\sim$3.65, 3.69, 3.71, 3.79, 3.85, 3.95, 3.98, and 4.01 \micron. As illustrated in Figure \ref{allch4}, these absorption features all correspond in wavelength with absorption bands of CH$_4$. The cross sections of methane (T=400 K, P=1 bar) are shown at the top of the figure at two resolutions ($\sim$1500 and 300); the locations of absorption features in WISE 0855's spectrum are shown as dashed lines to guide the eye. The wavelengths of these absorption features show good agreement with peaks in the methane cross sections. We therefore conclude that the features we observe in WISE 0855's L band spectrum are predominantly CH$_4$ features, in contrast to the M band spectrum published in \citet{Skemer16} which shows predominantly H$_2$O features.

We then compare the L band spectrum we observe in this work to other brown dwarfs and to Jupiter in Figure \ref{sequence}. We include, from hottest to coldest, SDSS 0857+57 (spectral type L8), SDSS 0758+32 (spectral type T2), and UGPS 0722-05 (spectral type T9). The T9 dwarf UGPS 0722-05 was the coldest brown dwarf for which a spectrum in L band was available previous to this work. The spectra of SDSS 0857+57 and SDSS 0758+32 were taken with the Near InfraRed Imager and spectrograph \citep[NIRI; ][]{Hodapp03} and published in \citet{Stephens09}. The spectrum of UGPS 0722-05 was taken with the Infrared Camera and Spectrograph \citep{[IRCS;][] Kobayashi00} on the Subaru telescope and published in \citep{Leggett12}. The Jupiter spectrum was observed using the Short Wave Spectrometer (SWS) on the Infrared Space Observatory \citep{Irwin09}. 

 Methane absorption is clearly seen in the L8 and T2 spectra at 3.3--3.35 \micron, strengthening in amplitude with later spectral type. The slope of the overall spectrum changes from relatively flat to significantly more sloped with decreasing \teff. For the T9 and for WISE 0855, methane absorption is even stronger and the objects are not detected at high significance within the 3.3 \micron\ feature, but other weaker absorption features are seen from 3.6--4.1 \micron\ (see Figure \ref{allch4}); these features increase in strength with decreasing \teff. Jupiter's L band spectrum also shows a number of the same absorption features. However, Jupiter's spectrum in L band is dominated by reflected light rather than thermal emission, so we expect the overall shape of the spectrum and amplitude of the features to be distinct from the brown dwarfs, which have contributions only from thermal emission. 

Model pressure--temperature profiles with effective temperatures and gravities that correspond to the objects shown in the top panel of Figure \ref{sequence} are shown in the bottom panel of that figure (models from 500--1200 K are from Marley et al., in prep.). The measured pressure--temperature profile of Jupiter from \citet{Seiff98} is also shown. The approximate locations of the L band photospheres are shown as a thick line. The dominant carbon and nitrogen chemistry, assuming chemical equilibrium, are shown as dash-dot lines. This chemistry is strongly temperature dependent: for the warmest model (\teff=1200 K), the atmosphere is expected to have more CO than CH$_4$ throughout. For the next model (\teff=1100 K), the top of the atmosphere has more CH$_4$ while the bottom has more CO. The three coldest atmospheres are dominated by CH$_4$ throughout. The coldest model (\teff=250 K) and Jupiter have much more NH$_3$ than N$_2$.

\subsection{How Different Atmospheric Properties Change Model Spectra}

\begin{figure*}[t]
\center \includegraphics[width=5.5in]{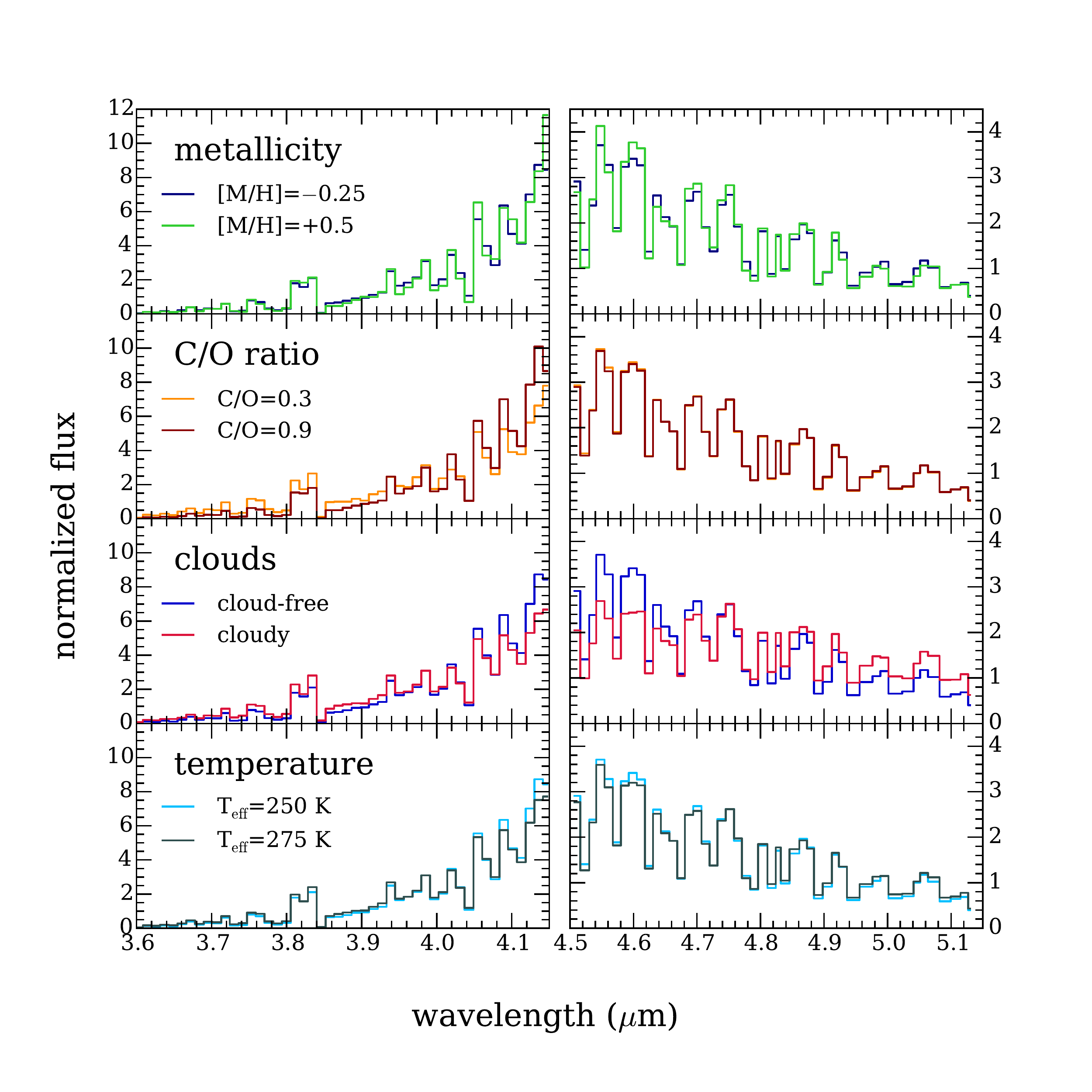}
 \caption{Model spectra in L and M bands illustrating the effect of changing four different model parameters. From top to bottom, these panels show models with different metallicity, C/O ratio, clouds, and temperature. Models are the same as in Figure \ref{model_spectra}. All models have surface gravity of 100 m/s$^2$. Top panel: 250 K, cloud-free, C/O ratio of 0.6, and [M/H] of -0.25 and +0.5. 2nd panel: 250 K, cloud-free, [M/H] of -0.25, and C/O ratio of 0.3 and 0.9. 3rd panel: 250 K, C/O ratio of 0.6, [M/H] of -0.25, and cloud-free and cloud parameter $f_{\rm sed}=4$. Bottom panel: cloud-free, [M/H] of -0.25, C/O ratio of 0.6, and temperature of 250 K and 275 K. }
\label{model_spectra_binned}
\end{figure*}

\begin{figure*}[t]
\center \includegraphics[width=5.5in]{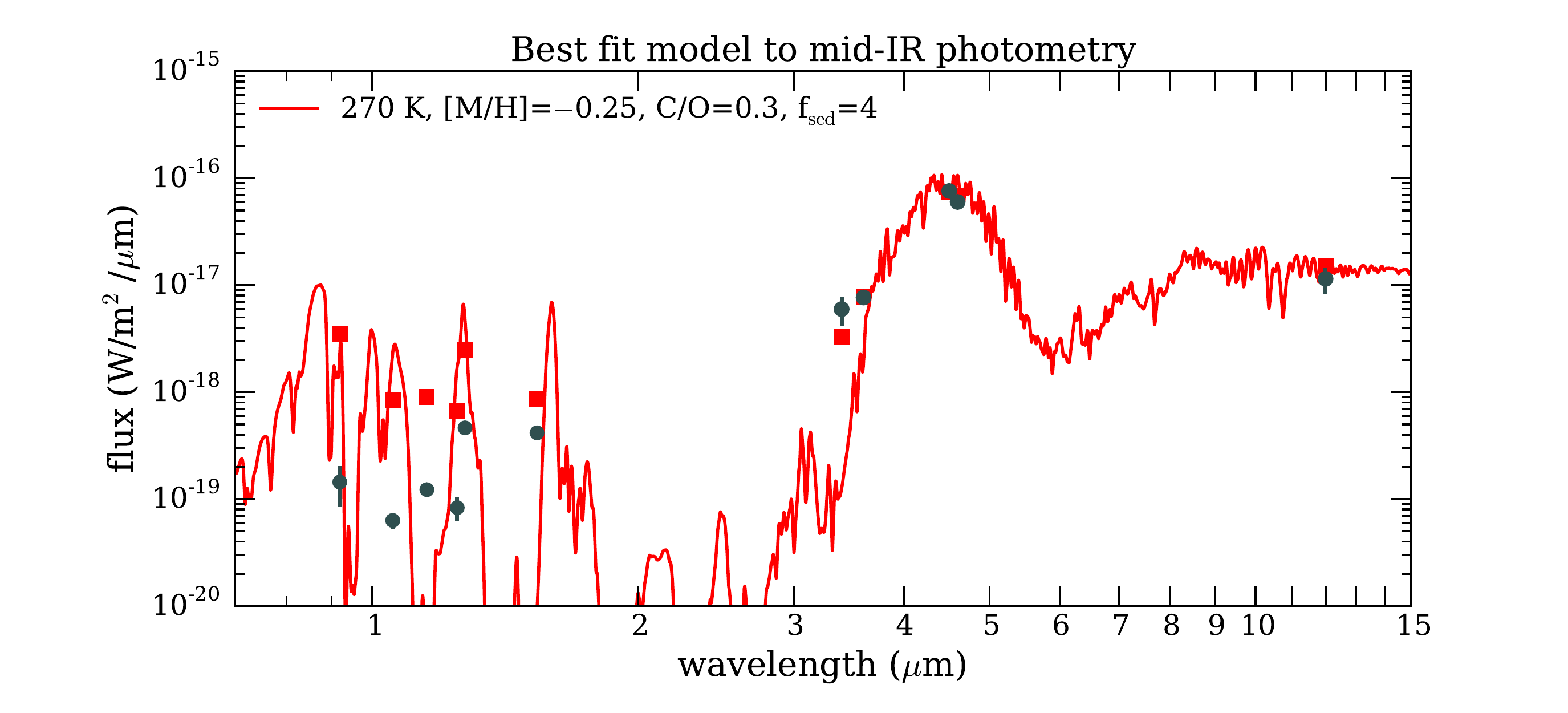}
\vspace{-5mm}
\center \includegraphics[width=5.5in]{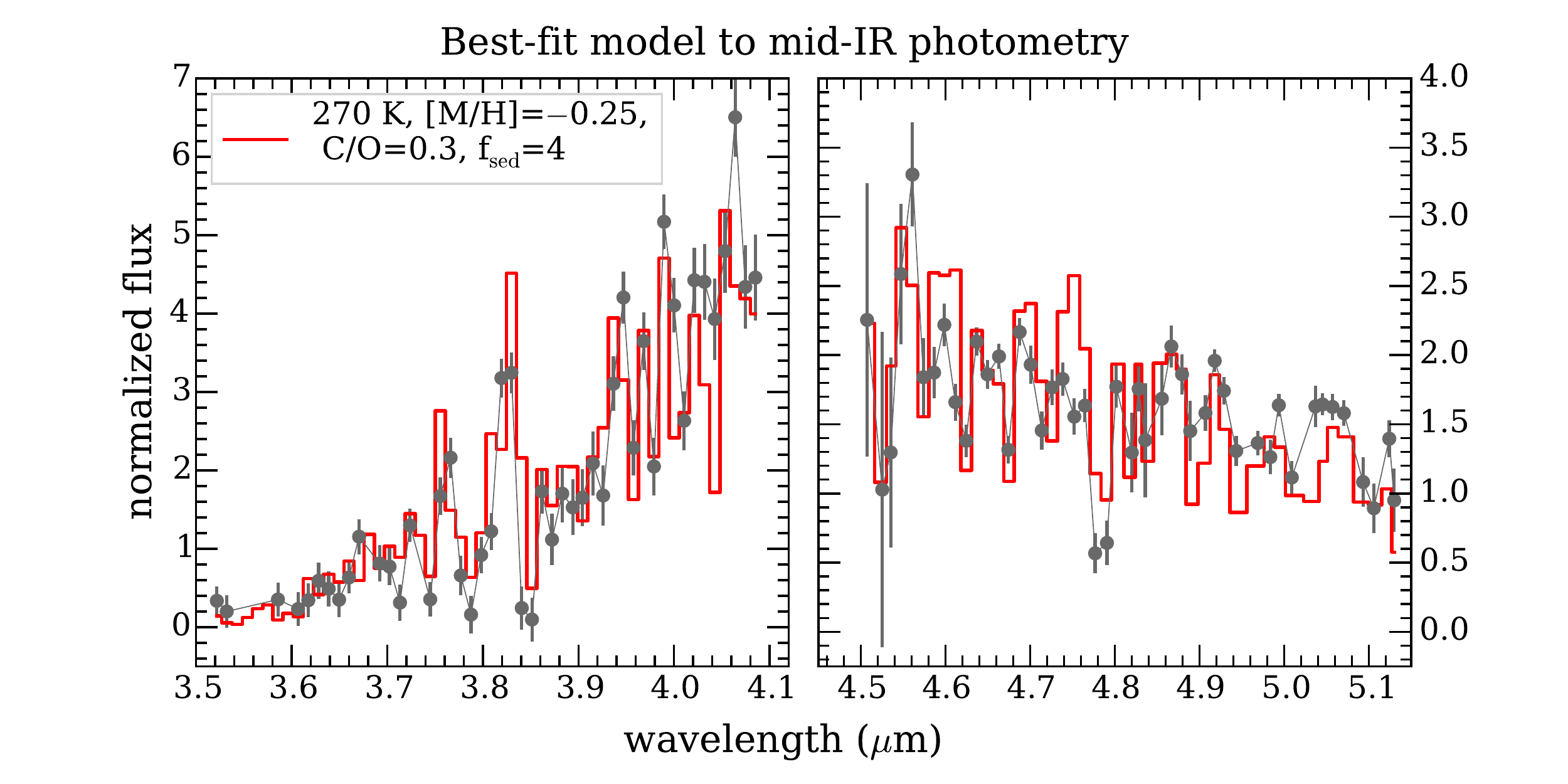}
 \caption{The best-fitting model to the mid-infrared photometry. The top panel shows the spectrum from 0.7 to 15 \micron. Data points are shown as gray points with error bars. Model photometry is shown as filled red squares. The bottom panel shows the L band (left) and M band (right) spectra. The data is shown as gray points with error bars, and the model spectrum is binned to the same wavelength grid and shown as a red line. The model matches the mid-infrared photometry, but is substantially brighter than the observations at near-infrared wavelengths.  }
\label{bestfit_midIRphot}
\end{figure*}

\begin{figure*}[t]
\center \includegraphics[width=5.5in]{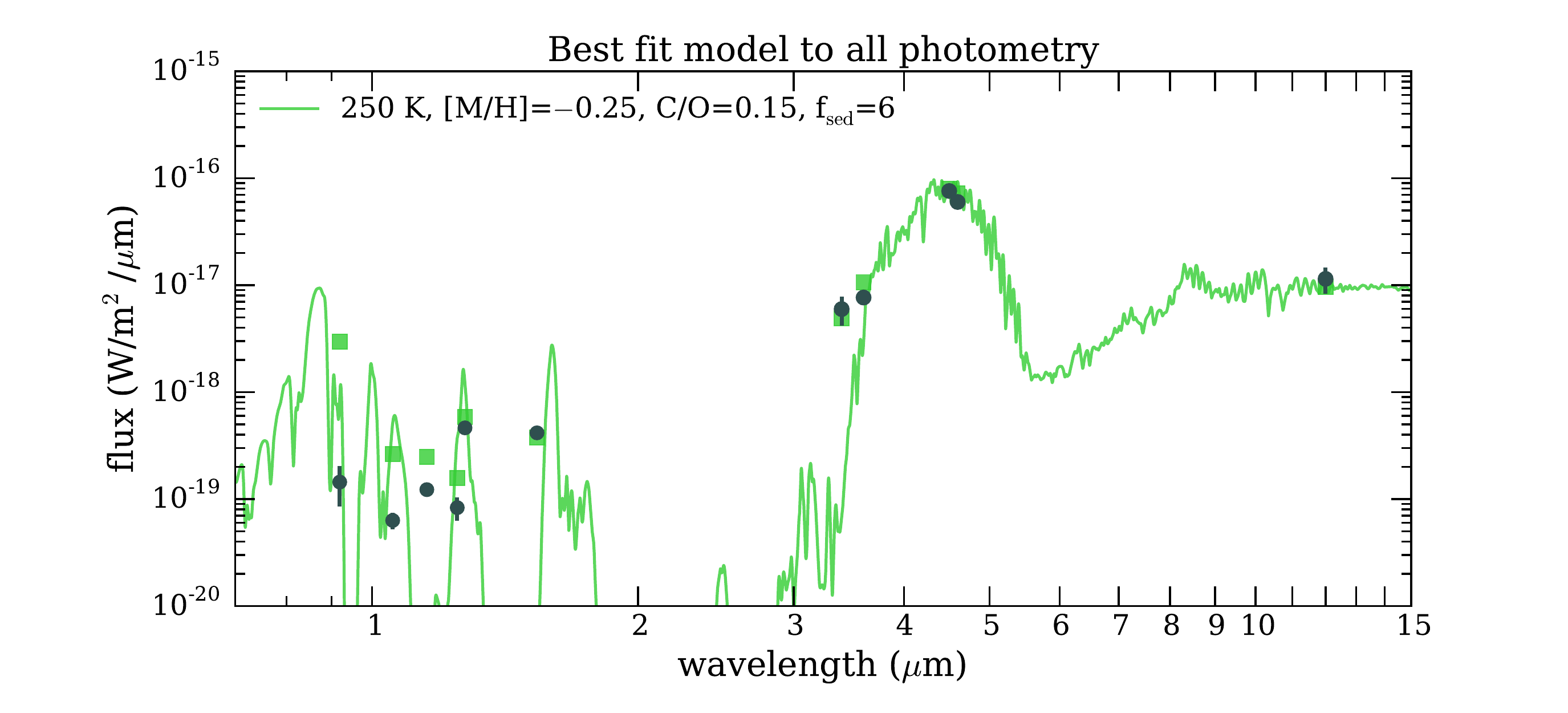}
\vspace{-5mm}
\center \includegraphics[width=5.5in]{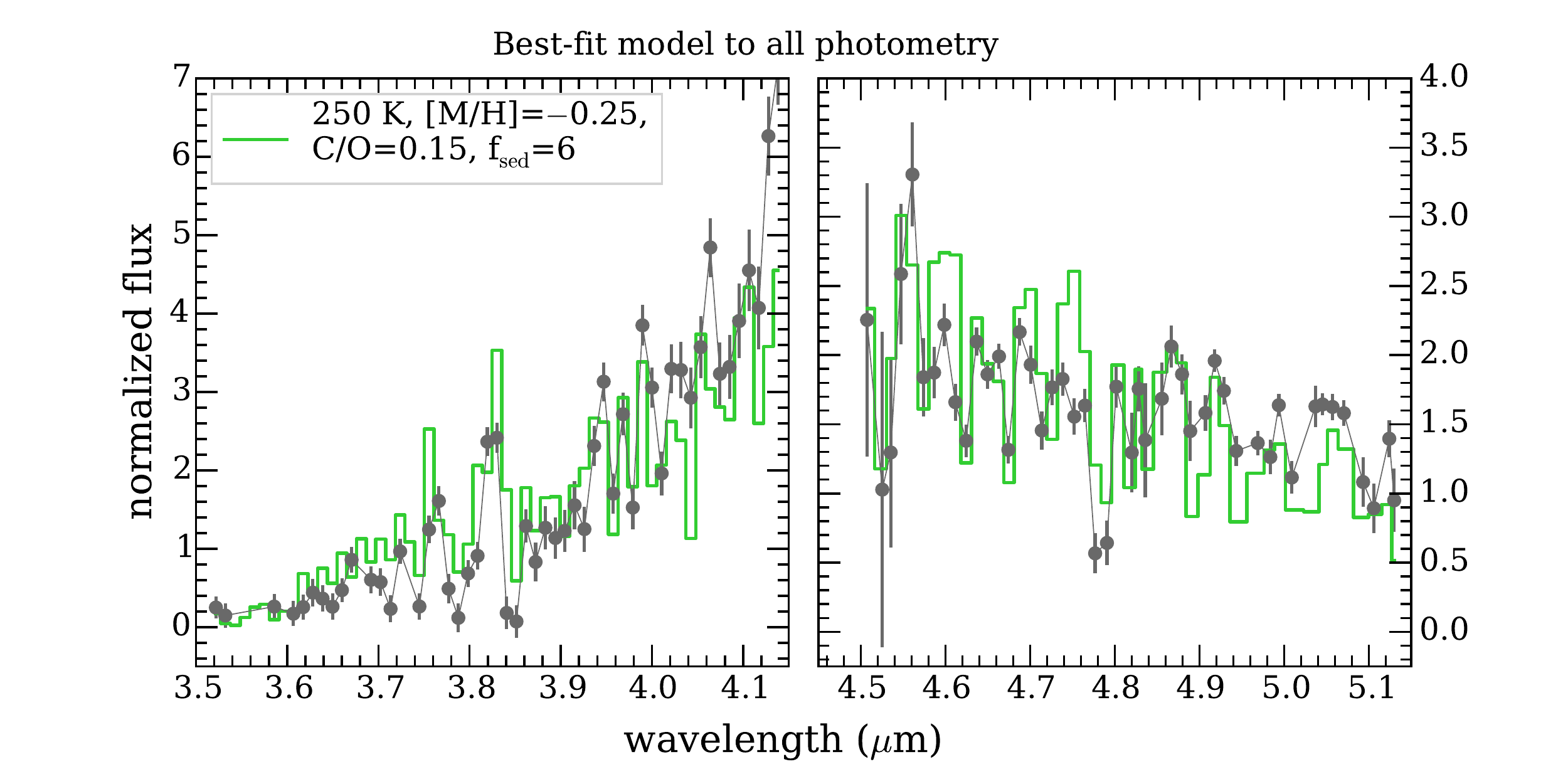}
 \caption{The best-fitting model to the near- and mid-infrared photometry. The top panel shows the spectrum from 0.7 to 15 \micron. Data points are shown as gray points with error bars. Model photometry is shown as filled green squares. The bottom panel shows the L band (left) and M band (right) spectra. The data is shown as gray points with error bars, and the model spectrum is binned to the same wavelength grid and shown as a green line. The model is substantially brighter than the observations at the bluest near-infrared wavelengths.  }
\label{bestfit_nearIRphot}
\end{figure*}

The four physical properties of the atmosphere that we focus on in this study are metallicity, C/O ratio, clouds, and effective temperature. Each of these properties has a significant effect on the resulting thermal emission spectrum. These effects are shown for a selection of illustrative models in Figures \ref{model_spectra} and \ref{model_spectra_binned}; in Figure \ref{model_spectra}, the full spectrum from 0.8--15 \micron\ is shown, and in Figure \ref{model_spectra_binned}, the spectrum is binned to approximately the resolution of the observed L and M band spectra and normalized. 

Lower metallicity models typically have broader near-infrared spectral features and more flux at the K band peak ($\sim$2.1 \micron). Because lower metallicity models have less CH$_4$, the emergent flux from 3--4 \micron\ increases at lower metallicity (see, e.g., Tremblin et al. models in \citet{Leggett17}, their Figure 8). As discussed in Section \ref{prev_photom}, a long-standing discrepency between model Y dwarfs and observed Y dwarfs is that the models are too faint in bandpasses that probe this 3--4 \micron\ flux, so in general we find that lower metallicity models match the mid-infrared photometry better than higher metallicity models. Decreasing the metallicity subtly decreases the size of features in L band, while making the overall slope of the M band spectrum shallower. 

Decreasing the C/O ratio is another way to decrease the overall methane abundance in our models, since methane is the dominant carbon-bearing species in the atmosphere. For these models, we hold oxygen abundance steady while changing the carbon abundance to change the C/O ratio. Lower C/O ratio model spectra have more flux within major CH$_4$ features including the red side of H band (1.65 \micron), 3--4 \micron, and 7--8 \micron. Decreasing the C/O ratio makes the L band slope more shallow, while leaving the M band spectrum essentially unchanged. 

The formation of water clouds has a substantial effect on the spectrum. Clouds both provide a continuum opacity source in the atmosphere and make the atmosphere slightly warmer, especially in the upper regions where the clouds are forming. Water clouds are especially interesting because they have strongly wavelength-dependent opacity (see discussion in Section \ref{h2o_clouds}), with more strongly absorbing optical properties in the thermal infrared than in the near-infrared. The effect of water ice clouds on the spectrum is apparent in Figure \ref{model_spectra}, where the cloud-free and cloudy models have distinctly different SEDs; the cloudy model has a lower peak flux from 4--5 \micron, and higher flux in absorption bands and through the rest of the spectrum. In Figure \ref{model_spectra_binned}, clouds change both the L and M band spectra; the cloudy L band spectrum has a shallower slope, while the cloudy M band spectrum is flatter with smaller absorption features. 

The bottom panels of Figures \ref{model_spectra} and \ref{model_spectra_binned} show the effect of changing the temperature; in general, changing the temperature does not change the shape of the spectrum, but just scales the flux. The lack of large changes in spectral shape is because the chemistry does not change substantially across this small temperature range.

\section{Matching Model Spectra with Observed Spectra \& Photometry}

\begin{figure*}[t]
\center \includegraphics[width=5.5in]{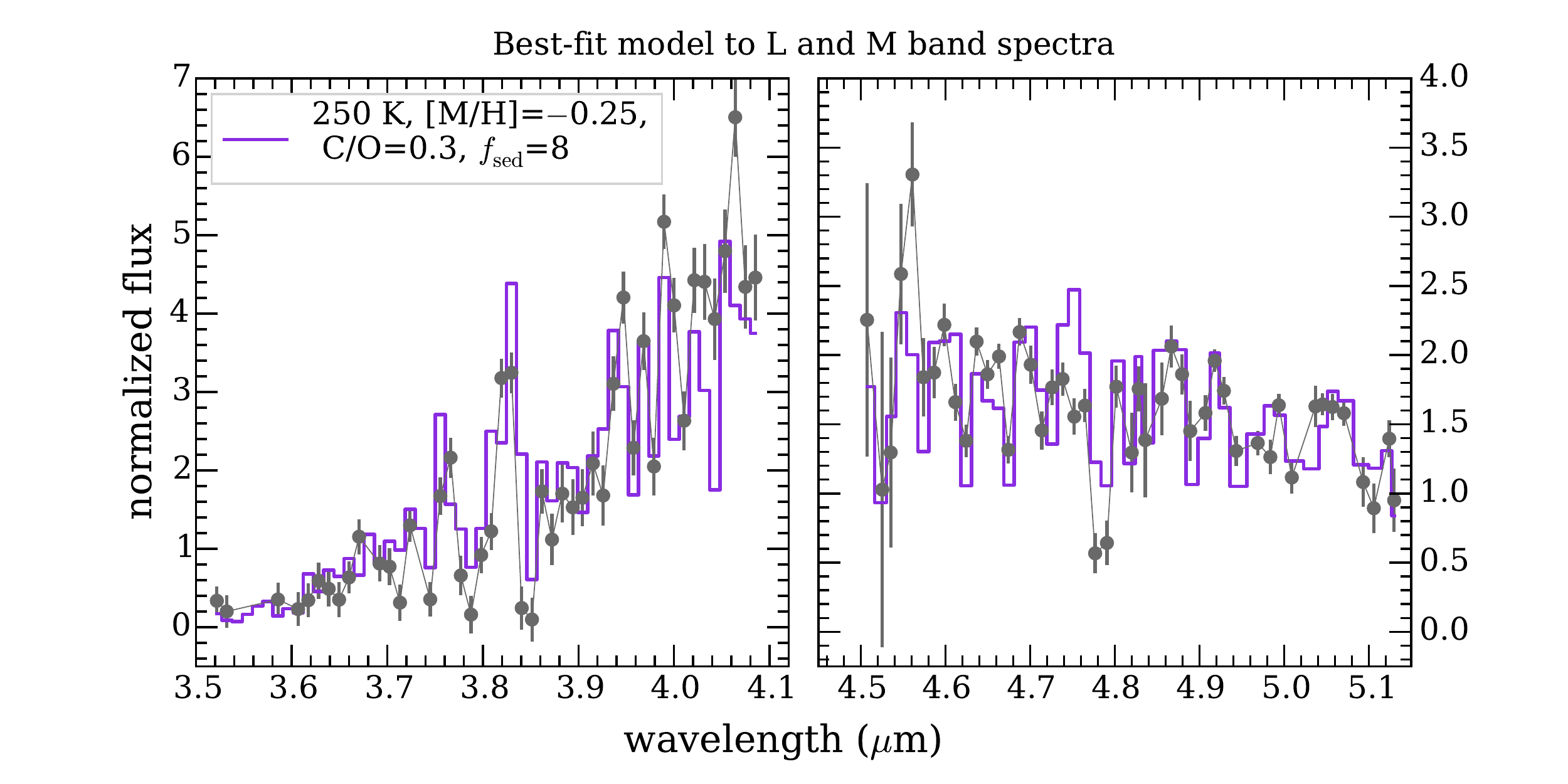}
 \caption{Best-fitting model spectra to the observed L and M band spectra. L band is shown on the left and M band is on the right. The data are shown as gray points with error bars, and the model spectrum is binned to the same wavelength grid and shown as a purple line.  }
\label{bestfit_spectra}
\end{figure*}

There is now a sufficient quantity of high quality data for WISE 0855 that it is a challenge to find models that fit the full dataset. We do not set out to perform a quantitative numerical analysis of the full range of models that provide the best-fits to the spectra and photometry of this object. We suggest that this endeavour is better suited to future studies using retrieval techniques to fit this wealth of data, as developed for brown dwarf studies by \citet{Line15, Line17} and \citet{Burningham17}. Instead we outline the interesting atmospheric physics that the current dataset suggests is important, and which will be tested in detail with upcoming observations. 

We locate a set of illustrative models that provide relatively good matches to portions of the spectra and photometry of WISE 0855. Using those models, we suggest that evidence exists for (1) relatively low methane abundance, (2) water ice clouds, (3) a dearth of PH$_3$ compared to Jupiter, and (4) a deep continuum opacity source.  

\subsection{Less Methane Than Expected?}

Our best-fit models to the available data for WISE 0855 are shown in Figures \ref{bestfit_midIRphot}, \ref{bestfit_nearIRphot}, and \ref{bestfit_spectra}. Figure \ref{bestfit_midIRphot} shows the data with a model from our small grid that best-matches the mid-infrared photometry. We include the three bluest WISE bands (3.4, 4.6, and 12 \micron) and two Spitzer IRAC bands (3.6 and 4.5 \micron). Figure \ref{bestfit_nearIRphot} shows the data with a model from our small grid that best-matches the full set of photometry from \emph{HST}, \emph{Spitzer}, and \emph{WISE}. Figure \ref{bestfit_spectra} shows the spectral data with a model that best matches the L and M band spectra.

We find that solar composition, solar carbon-abundance models fit the mid-infrared color very poorly, in agreement with a variety of previous studies \citep[e.g.,][]{Luhman14, Leggett17}. We start by focusing on finding a family of models that can match these mid-infrared data, since >99\% of the thermal emission is expected to be emitted at wavelengths longer than 2 \micron.  

As shown in Figure \ref{model_spectra}, there are a variety of ways to increase the flux from 3--4 \micron\ relative to 4--5 \micron\ as is needed to fit the observed photometry: decrease the metallicity, decrease the C/O ratio, and increase the cloud opacity. A variety of models are found to provide reasonable fits to the data, including some at extremely low metallicities ([M/H]=$-0.75$ to $-1$), well outside of the expected range of metallicities for stars in the local neighborhood \citep{Hinkel14}. If we stay within more conservative bounds of plausible solar neighborhood metallicities ([M/H]=$-0.25$ to $1.0$), we find that models with [M/H]=$-0.25$, C/O=0.3 (half solar C/O), and with water ice clouds provide the best matches to the mid-infrared SED of WISE 0855. This particular model has goodness-of-fit metric $\chi^2/N$ of 3.0 compared to the observed mid-infrared photometry, where $\chi^2$ is $\sum_{i=1}^N\frac{(\rm{data_i}-\rm{model_i})^2}{\sigma_i^2}$ and $N$ is the number of data points. We find similarly good fits ($\chi^2/N\sim$4--5) for models with solar metallicty and even-more-sub-solar C/O (0.15, quarter-solar C/O). These low-methane abundance models are substantially better matches to the observed mid-infrared photometry than the best-fit solar metallicity and solar C/O ratio models, which have a goodness-of-fit $\chi^2/N>$41.

The bottom panel of Figure \ref{bestfit_midIRphot} shows the same model zoomed in to compare to the L and M band spectra of WISE 0855. The model has goodness-of-fit $\chi^2/N$=6.7 (L band) and $\chi^2/N$=7.6 (M band). 

Fitting to the full set of photometry of WISE 0855 (Figure \ref{bestfit_nearIRphot}) provides qualitatively similar results. We favor somewhat colder models (\teff=250 instead of 270 K), but similarly find lower metallicities and lower C/O ratios provide the best matches. We note that none of the models are good matches to the full dataset, with a best goodness-of-fit in the grid $\chi^2/N>$ of 23.7. We suggest a possible mechanism that could explain this discrepancy in Section \ref{deepcloud}. 

In general, these results suggest that a promising way to match the infrared colors of WISE 0855 is to decrease the amount of methane in the atmosphere, either through primordial abundances (low metallicity or low C/O ratio) or by an as-of-yet-unaccounted for sink of carbon or methane. 

We do not however expect mixing from deep, hot layers of the atmosphere to be an effective mechanism for decreasing the CH$_4$ abundance: as shown in Figure \ref{sequence}, for a $\sim$250 K object, even deep layers at nearly 1000 bar pressures are dominated by CH$_4$. Using more detailed models including vertical mixing, \citet{Morley14a} (their Figure 12) similarly find no significant spectral changes due to disequilibrium carbon (and nitrogen) chemistry from vertical mixing for the coldest objects.  

As discussed previously, the [3.6]--[4.5] color is discrepant between models and observations for all T and Y dwarfs cooler than 600 K; the proposed solution of lower methane absorption may therefore potentially solve this problem for WISE 0855 and other cold dwarfs. While a single object like WISE 0855 might have a primordial composition that deviates from solar, it is unlikely that \emph{all} T and Y dwarfs have sub-solar compositions. \citet{Leggett17} investigated metallicity variations in the population of Y dwarfs and did not find that WISE 0855 appeared to be an outlier in its composition. Given that a systematic low primordial metallicity or C/O ratio is unlikely to be causing these differences, further modeling work is needed to determine the root cause of the apparent low methane absorption in the population of T and Y dwarfs studied to date.

\subsubsection{Can Departures from Radiative--Convective Equilibrium Reproduce WISE 0855's Spectrum?}

\begin{figure*}[t]
 \includegraphics[width=7.5in]{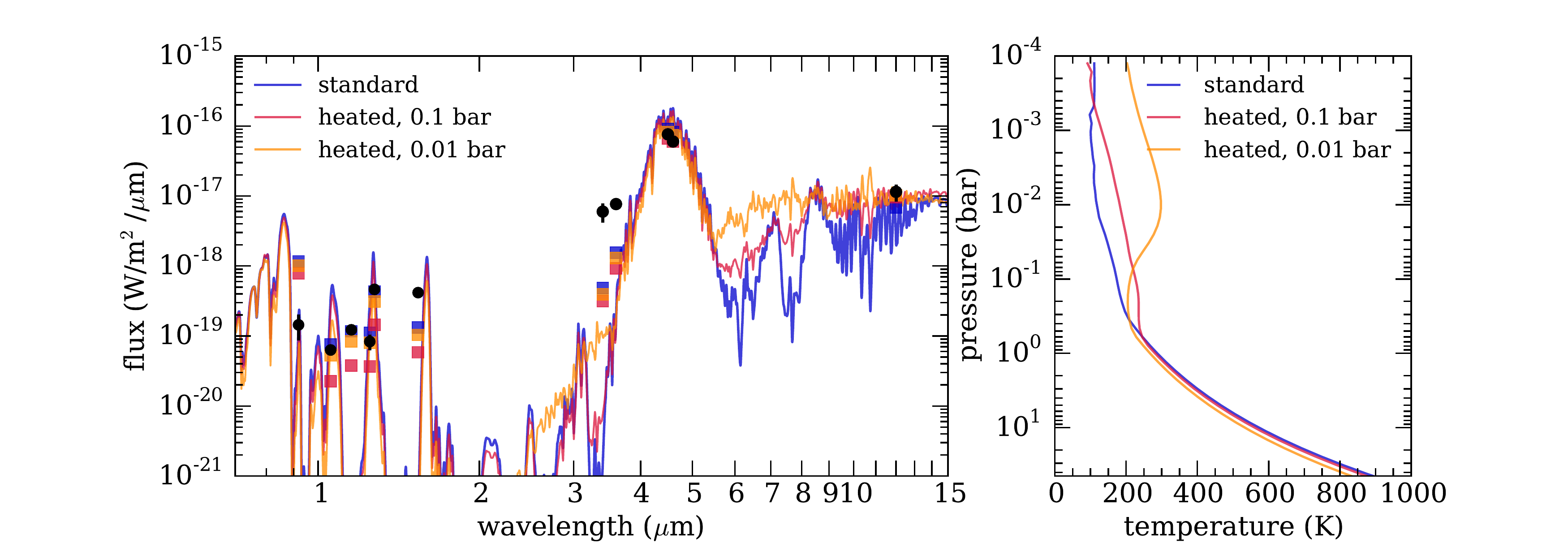}
 \caption{Spectra and pressure--temperature profiles of models with heated upper atmospheres. The left panel shows model spectra compared to the observed photometry of WISE 0855. The blue model shows a standard model (250 K, [M/H]=0.0, log g=4.0, C/O=0.6, cloud-free). The red and orange models show models with approximately the same emergent flux as the standard model (\teff=256 K) but with heated upper atmospheres (heating centered at 0.1 and 0.01 bar respectively). The right panel shows the pressure--temperature profiles for the same three models.  }
\label{heatedupper}
\end{figure*}

We explore one such avenue for decreasing methane \emph{absorption} without changing the methane \emph{abundance}: processes that drive departures from radiative--convective equilibrium. Previous observational studies have invoked heating in the radiative part of the atmosphere of a brown dwarf to match the mid-infrared colors of some brown dwarfs. For example, \citet{Sorahana14} and \citet{Burningham17} found that the spectra of several L dwarfs can be better matched with atmospheres that depart from radiative equilibrium. 

Here we run several models to test how atmospheric heating between 0.1 and 100 mbar could change the spectra of a 250 K object like WISE 0855. We incorporate atmospheric heating following the approach of \citet{Marley99} and \citet{Morley14b} by adding energy at specified pressure levels of model atmospheres as we calculate the pressure--temperature structure in radiative--convective equilibrium. The perturbations have the shape of a Chapman function with a width of a single pressure scale height, which is often used to represent heating by incident flux within molecular bands \citep[e.g.,][]{ChambHunt}. We center the heating at 0.01 and 0.1 bar. We use an `internal temperature' of 225 K and add 10$^5$ erg cm$^{-2}$ s$^{-1}$, resulting in a \teff\ of 256 K (this results in a spectrum that is $\sim$1.7 times brighter than a ``standard''  225 K model). 

The resulting model spectra and pressure--temperature profiles are shown in Figure \ref{heatedupper}. We find that the shape of the mid-infrared spectrum changes considerably as the atmosphere is heated at both 0.1 and 0.01 bar. In particular, the deep absorption features at 3.3, 6, 8, and 10 \micron\ become less deep. However, the resulting model photometry at 3--4 \micron\ is not actually closer to the observed photometry; the effect is not large enough for atmospheric heating, at least of this magnitude, to resolve the discrepancies between the observed and model photometry.  

Deviations in the adiabatic structure of the deep atmosphere have also been suggested as a mechanism that may be important in brown dwarfs \citep{Tremblin15}. However, we do not model this here since \citet{Leggett17} explore this idea in more detail and find that a modified shallower adiabat does not bring the model photometry substantially closer to the observed photometry (e.g., their Figure 12). 

Future observations with \emph{JWST} would likely be sensitive to such major deviations from radiative--convective equilibrium. Further modeling work on the climate and atmospheric circulation of 250 K planets would provide theoretical insight into the causes of such deviations; for example, energy could be deposited by the breaking of upward propagating gravity waves in the radiative part of the atmosphere.

\subsection{Water Ice Clouds In WISE 0855}

\subsubsection{Previous Literature on Clouds in Y Dwarfs and WISE 0855}

Since the first modeling efforts for cold brown dwarfs and exoplanets, the role of water clouds has been considered. For example, \citet{Marley99} and \citet{Sudarsky00} modeled the effect of water clouds on the albedos of giant exoplanets; \citet{Burrows04} also considered water clouds in exoplanets. \citet{Sudarsky03}  and \citet{Sudarsky05} calculated the thermal emission of exoplanets that include water clouds, finding that they have a strong effect on the emergent spectrum. \citet{Burrows03b} and \citet{Morley14a} included water ice clouds in thermal emission models of cold brown dwarfs. 

Since the discovery of WISE 0855, a number of observational studies have addressed the role of water ice clouds in its atmosphere. \citet{Faherty14b} detected WISE 0855 in the near-infrared for the first time; they  compared the observed photometry to solar composition, chemical equilibrium models and found that the cloudy models (including water ice and sulfide clouds) provided a better fit to the J3$-$[4.5] color than cloud-free models. \citet{Luhman16} considered the role of disequilibrium chemistry and found that this provided an alternative explanation for the J$-$[4.5] color; both \citet{Luhman16} and \citet{Schneider16} found that no models provided adequate fits to all photometry. \citet{Skemer16} published the first spectrum of WISE 0855 in M band, noting the presence of water vapor features. They found that the amplitudes of those water vapor features appear muted, which could be better reproduced with models including simple gray cloud opacity at a height in the atmosphere that could be associated with a water ice cloud than with cloud-free models. \citet{Esplin16} published the first photometric monitoring of WISE 0855, finding that small deviations in cloud-covering fraction on each hemisphere could reproduce the observed $\sim$4\% variability in the mid-infrared. That study also claimed that the simplifications in the modeling scheme of \citet{Skemer16} -- in particular the use of gray cloud opacity when water ice is very non-gray, and the use of non-self-consistent clouds in the atmosphere -- may have affected the conclusions of \citet{Skemer16}. 

In this section, we build upon the previous work both theoretical and observational, to further investigate the possibility of water ice clouds in the atmospheres of cold objects like WISE 0855.

\subsubsection{The Spectral Effect of Non-gray Water Clouds} \label{h2o_clouds}

\begin{figure}[t]
\center \includegraphics[width=3.5in]{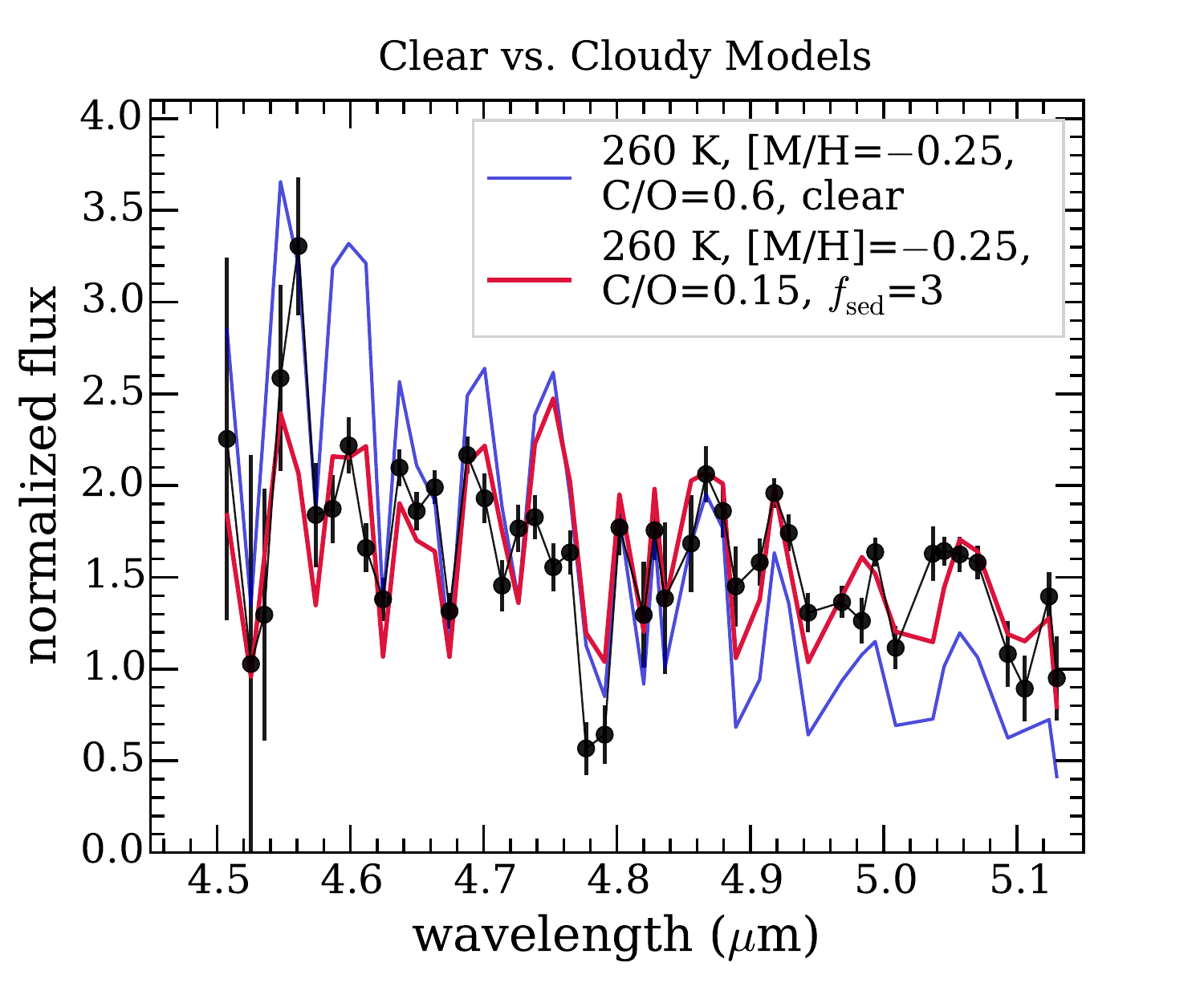}
 \caption{The best-matching cloudy and clear model spectra compared to the data in M band. The best-matching cloudy model is shown in red, and the best-matching clear model is shown in blue. }
\label{cloudsvsclear}
\end{figure}

The shape of WISE 0855's mid-infrared spectrum holds more information than the photometry alone. We have compared the overall slope of the L and M band spectra and the size of spectral features in each bandpass. We find that only models that include clouds can fit the flat slope of the M band spectrum and muted size of the features. Figure \ref{bestfit_spectra} shows a model fit to both L and M band data together ($\chi^2/N$=7.2 in L band, $\chi^2/N$=5.1 in M band). For comparison, the nominal best-fit model to the L band data alone has a goodness-of-fit $\chi^2/N$=5.2; the nominal best-fit model to the M band data has a goodness-of-fit $\chi^2/N$=5.0.

The shape of the M band data provides more information about water clouds with fewer degeneracies than the L band data; for example, C/O ratio and clouds have a similar effect on the shape of the L band spectrum at this resolution, but a very different effect on M band (see Figure \ref{model_spectra_binned}). Out of the grid of models considered here, the cloud-free models provided the worst fits in M band, with $\chi^2/N$>17. 

Figure \ref{cloudsvsclear} shows a typical cloudy model and cloud-free model compared to the spectroscopic data. In this figure, the best-fitting cloud-free model has goodness-of-fit $\chi^2/N=$17.4, while the cloudy model has a goodness-of-fit $\chi^2/N=$5.0. We also note that the spectrum of WISE 0855 should be sensitive to clouds at longer wavelengths, including those probed by the WISE W3 photometry. We find that the cloud-free models do tend to be fainter than the observed photometry, but not at high significance ($\sim1.1$--$2.4\sigma$). Cloudy models provide good matches to the W3 photometry. Future spectroscopic observations at these wavelengths will be informative. 

The necessity of clouds to match WISE 0855's spectrum is in qualitative agreement with the results from \citet{Skemer16}, in which simplified clouds were included in a cloud-free model atmosphere during post-processing when generating a model spectrum. Here we include `real' water ice clouds with the spectral properties of pure ice spheres, which are calculated self-consistently with the pressure--temperature profile in radiative--convective equilibrium. 

These advancements in the models are critical for the current work, which investigates both the spectra and photometry of WISE 0855. While the the nominal best-matching cloudy model with \emph{ad hoc} post-processed gray cloud opacity in \citet{Skemer16} provides a relatively good fit to the observed M band spectrum ($\chi^2/N\sim$4.4), it does not fit the observed photometry ($\chi^2/N\sim$1100). The limitations of the models used in \citet{Skemer16} make it hard to interpret the solidness of the conclusions regarding evidence for clouds; adding a continuum opacity source at the appropriate pressure level allowed the spectrum to approximately match, but the model itself became much too cold to fit the photometry since the temperature profile was not self-consistently calculated. In addition, gray opacities are not appropriate for non-gray water ice clouds; the effect of this wavelength-dependent opacity was not explored in \citet{Skemer16}. However, our new models, with self-consistent water ice clouds, are able to reproduce the features in M band as well as the post-processed clouds, without a negative effect to the SED fit. The models in this work therefore provide much stronger evidence that clouds are necessary to match the spectrum of WISE 0855. 

\begin{figure}[t]
\center \includegraphics[width=3.3in]{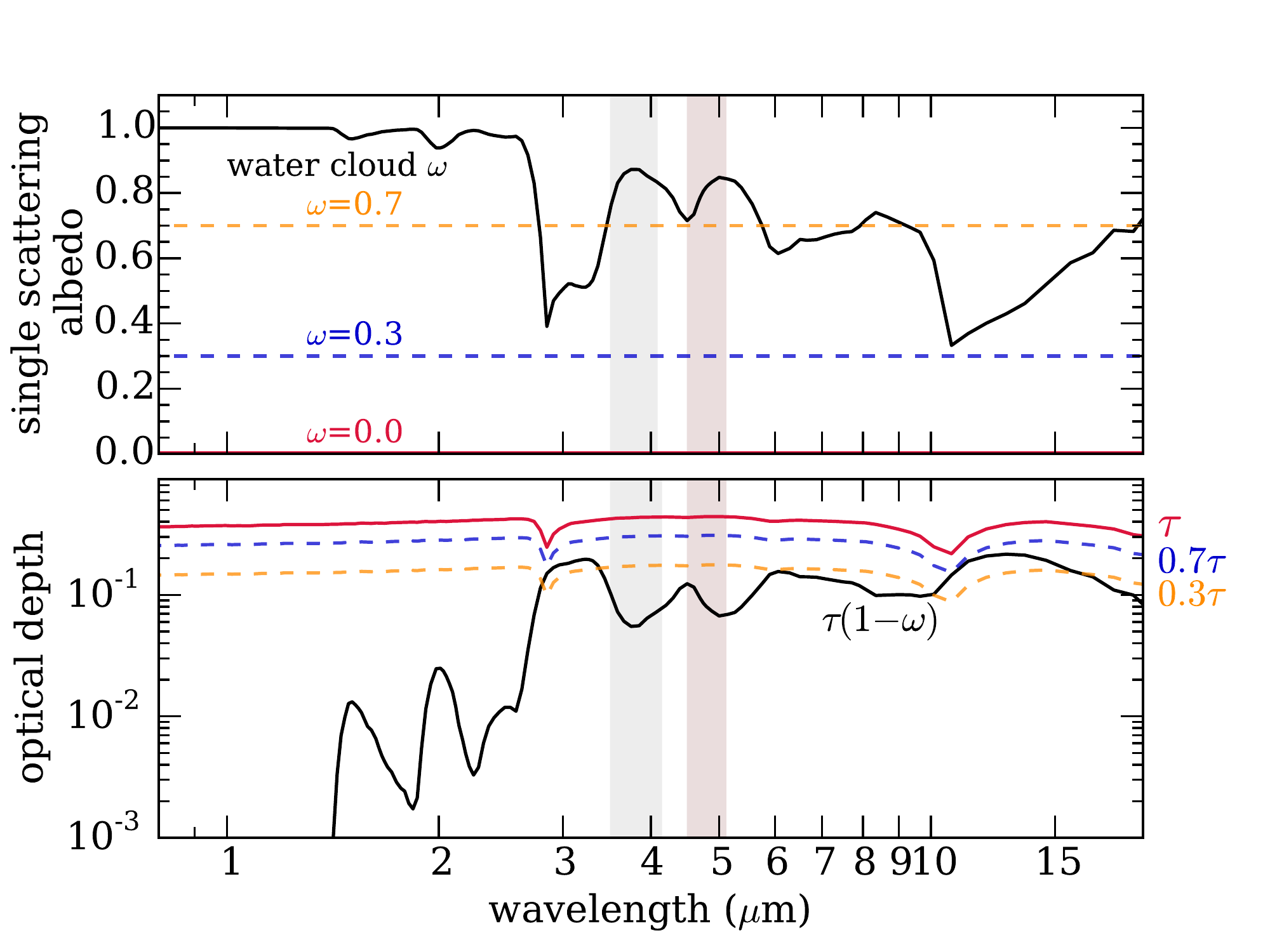}
 \caption{Optical properties of the water ice cloud. The top panel shows the single scattering albedo of a layer of water ice cloud. The model has \teff=250 K, \fsed=4, [M/H]=$-0.25$, C/O=0.6; the layer has P=0.05 bar and $T$=254 K. The constant single scattering albedos used in our test cases ($\omega$=0.0, 0.3, 0.7) are shown for reference. The bottom panel shows the cloud optical depth $\tau$. The black line shows the absorption component, $\tau(1-\omega)$. The red, blue, and orange lines show the same absorption component for the constant single scattering albedo test cases.  }
\label{water_properties}
\end{figure}

We find that since scattering by water ice particles is strongly wavelength-dependent, the assumptions about the cloud properties strongly impact the emergent spectrum. We show some simple tests to demonstrate this dependence, in which we change the single scattering albedo to be constant. Figure \ref{water_properties} shows the single scattering albedo of the water ice cloud: it scatters strongly at short wavelengths through the near-infrared, and then the single scattering albedo decreases sharply between 2 and 3 \micron\ and varies throughout the mid-infrared. The bottom panel of Figure \ref{water_properties} shows the optical depth $\tau$ of the cloud multiplied by (1$-\omega$) to show the absorptivity of the cloud. The black line shows true water ice, and the colored lines show the three test cases ($\omega=$0.0, 0.3, 0.7). 

\begin{figure}[t]
\center \includegraphics[width=3.7in]{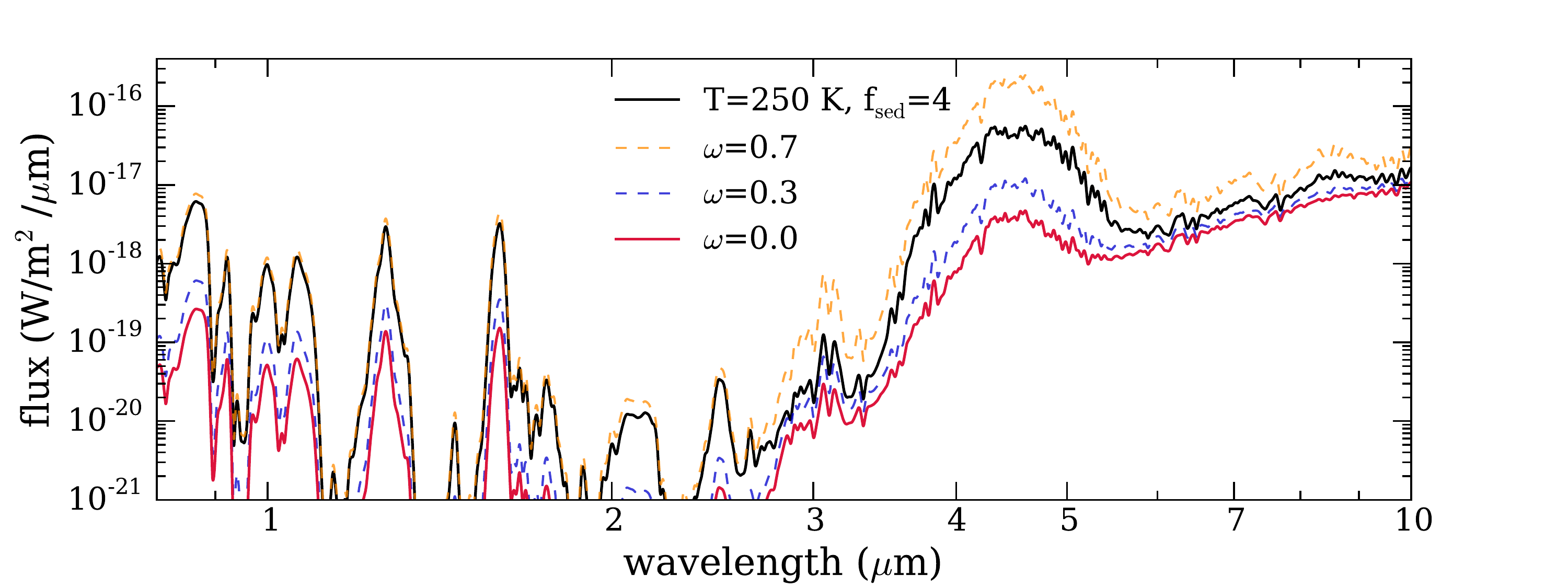}
\vspace{-0.4in}
\center \includegraphics[width=3.7in]{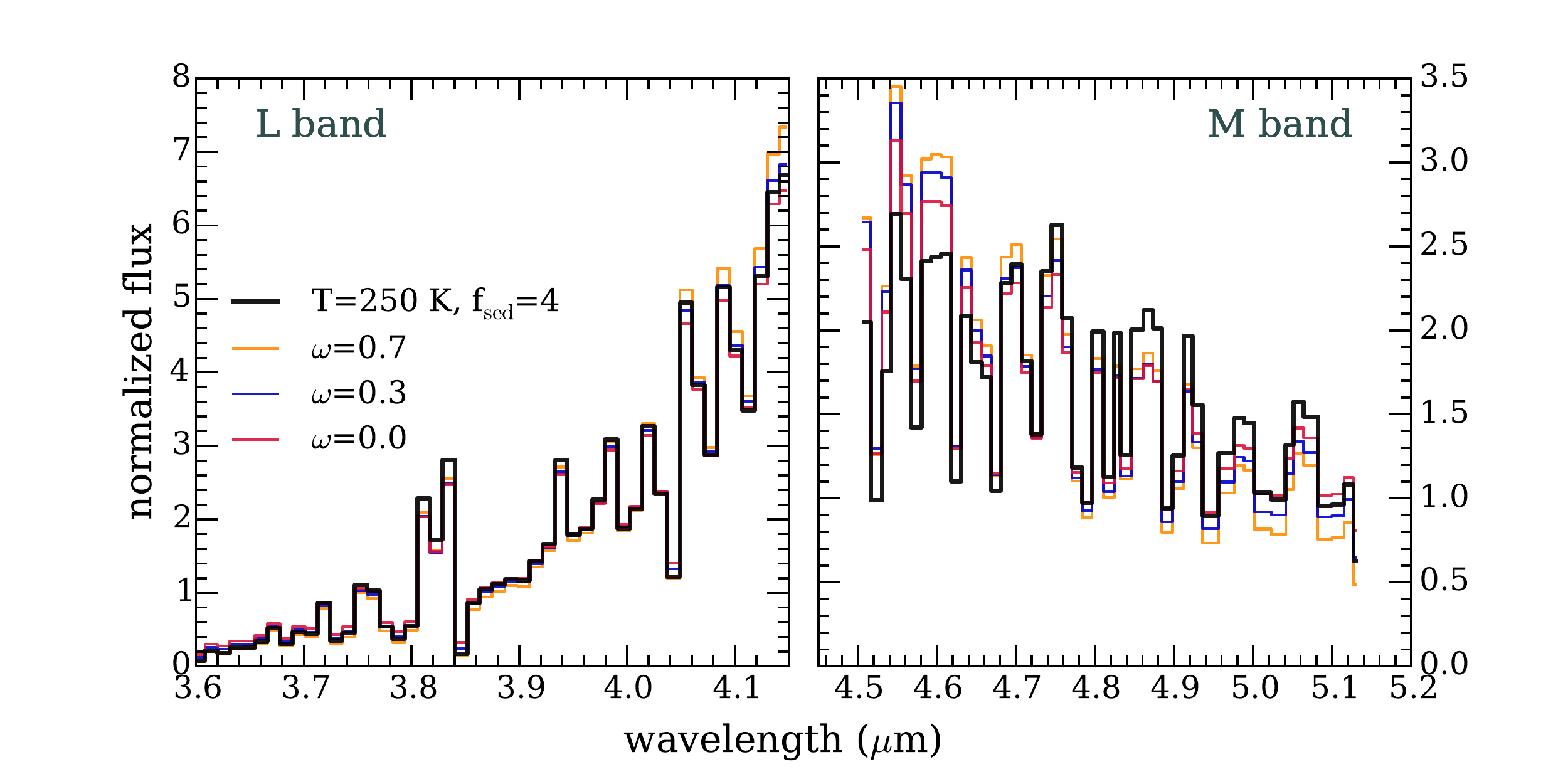}
 \caption{The effect of wavelength-dependent water ice clouds on model spectra. The top panel shows the spectrum from 0.8 to 10 \micron; the bottom panel bins the spectrum to the resolution of the observed L and M and spectra.}
\label{waterclouds_specs}
\end{figure}

The spectra of these test cases are shown in Figure \ref{waterclouds_specs}. The constant, highly scattering cloud ($\omega=0.7$) matches the water ice cloud at short wavelengths (<2 \micron) where the water cloud is strongly scattering, and is brighter than the water cloud model at longer wavelengths. In contrast, both more absorbing models ($\omega$=0.0 and 0.3) are fainter than the water cloud model at all wavelengths. However, zoomed in to L and M band in the bottom panel, all of the constant-$\omega$ test clouds have more strongly sloped M band spectrum than the water-ice cloud because of the change in the water ice single scattering albedo between 4.5 and 5 \micron. This test illustrates that the mid-infrared thermal emission spectrum of WISE 0855 is sensitive to the scattering properties of the cloud itself. Future work will be needed to study the effects of particle size, non-spherical particles, and impurities in the grains on the spectrum.

\subsection{The Apparent Lack of PH$_3$ in WISE 0855} \label{ph3}

\begin{figure}[thb]
\center \includegraphics[width=3.3in]{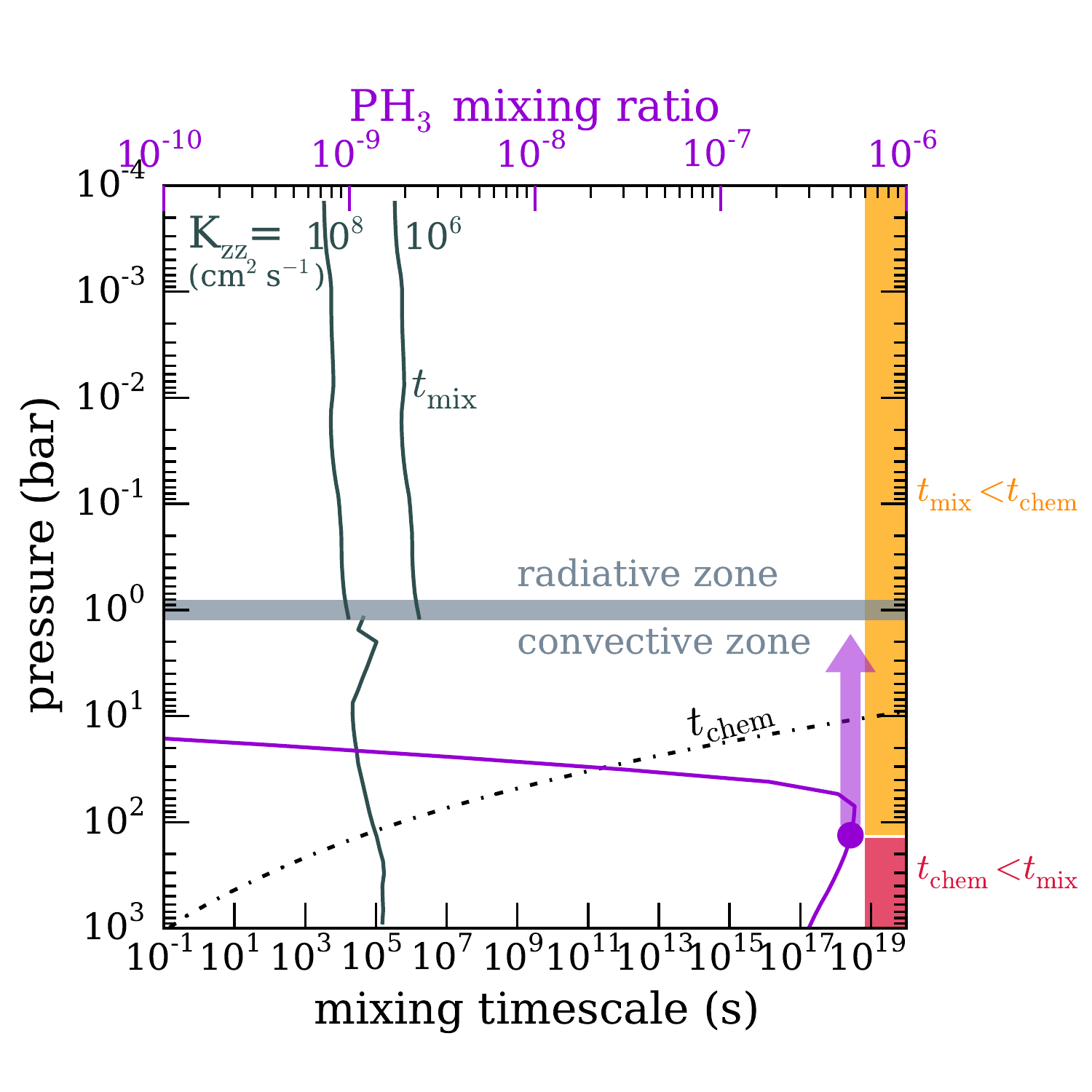}
 \caption{The mixing and chemical timescales in the atmosphere and the mixing ratio of PH$_3$. The bottom x-axis corresponds to timescales in the atmosphere; the mixing timescale is calculated using the eddy diffusion coefficient and scale height as described in the text, while the chemical timescale is the destruction timescale for PH$_3$ as calculated in \citet{Visscher06}. The top x-axis corresponds to the PH$_3$ mixing ratio assuming chemical equilibrium, shown in purple. The radiative--convective boundary is shown as a thick gray bar, with the radiative part of the atmosphere above and convective part below. In the convective zone, the mixing timescale is calculated using mixing length theory; in the radiative zone, the mixing timescale is calculated using two constant values of the eddy diffusion coefficient $K_{\rm zz}$. The right bar of the plot is shaded based on which timescale is faster. The purple arrow is drawn from the point where the chemical and mixing timescales are equal (the `quench' point). }
\label{ph3_kzz}
\end{figure}

\citet{Skemer16} found that while Jupiter has strong absorption features from PH$_3$ present in M band around 4.5--4.6 \micron, WISE 0855 did not show any evidence for these features. The PH$_3$ band is centered around 4.3 \micron\ (see Figure \ref{opacities_zoom}), between L and M bands, and extends into the red side of L band from 4.0--4.2 \micron. We see no evidence for PH$_3$ absorption in L band either, confirming the \citet{Skemer16} results. 

Figure \ref{ph3_kzz} shows the abundance of PH$_3$ expected assuming (rainout) chemical equilibrium, and the timescales for vertical mixing and chemical destruction of PH$_3$ in the atmosphere. The mixing ratio of PH$_3$ is as high as $\sim5\times10^{-7}$ around 100 bar and decreases sharply in abundance to be $<10^{-10}$ at 10 bar. In the convective region of the atmosphere, we calculate the approximate characteristic mixing timescale ($\tau_{\rm mix}\sim H^2/K_{\rm zz}$, where $H$ is the scale height and $K_{\rm zz}$ is the eddy diffusion coefficient) along the pressure-temperature profile using mixing length theory and find that it ranges between 10$^{4}$~s and 2$\times$10$^{5}$~s in the deep atmosphere. In the radiative part of the atmosphere we show the approximate mixing timescale corresponding to two constant $K_{\rm zz}$ values (10$^6$ and 10$^{8}$ cm$^2$s$^{-1}$). 

However, the timescale for PH$_3$ to be depleted (forming P$_4$O$_6$) is typically slower than the mixing timescales in cool planet and brown dwarf atmospheres \citep{Barshay78, Fegley94, Visscher06}. We follow the methods presented in \citet{Visscher06} to estimate PH$_3$ $\rightarrow$ P$_4$O$_6$ destruction timescales for the model in Figure \ref{ph3_kzz}: for the conditions here in the deep atmosphere (1200 K, 100 bar) the timescale is $\sim$10$^7$ s and increases quickly with decreasing temperature and pressure. The chemical timescale is faster than the mixing timescale in the deep atmosphere (P>130 bar) and slower than the mixing timescale in the rest of the atmosphere. This means that we expect the PH$_3$ abundance to be constant through the atmosphere above 100--200 bar, with a mixing ratio around 5$\times$10$^{-7}$.  

\begin{figure}[t]
\center \includegraphics[width=3.5in]{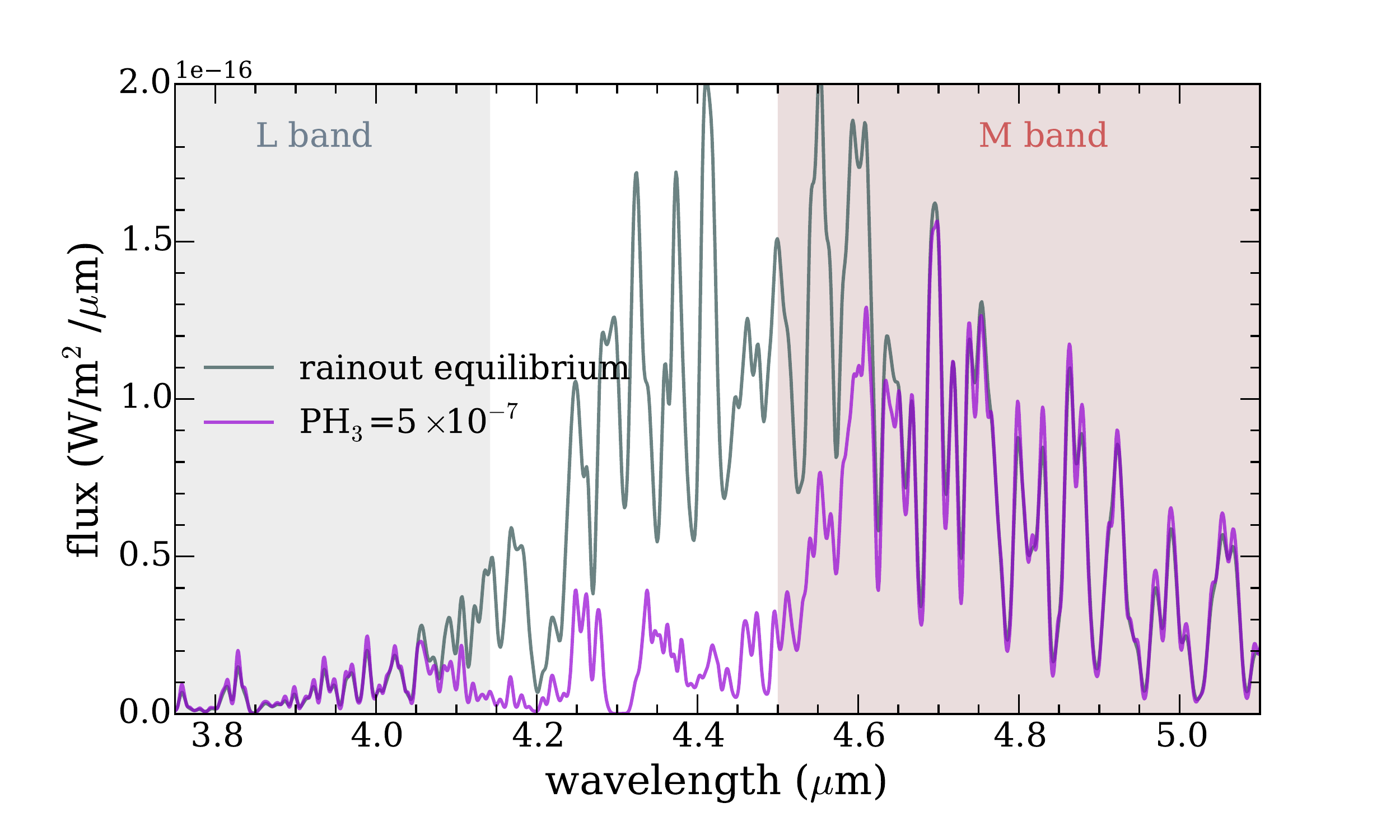}
 \caption{Spectra that show the effect of PH$_3$ on the spectrum. The gray line shows the spectrum assuming chemical equilibrium including rainout. The violet line shows the PH$_3$ abundance predicted for the calculated mixing and chemical timescales (see Figure \ref{ph3_kzz}). The wavelength regions of L and M bands are shaded.  }
\label{ph3_specs}
\end{figure}

Figure \ref{ph3_specs} shows how a typical model spectrum changes with additional PH$_3$ mixed from the deep atmosphere. The predicted mixing ratio of 5$\times10^{-7}$ results in a strong absorption feature from 4.05 to 4.6 \micron. Wavelengths within both L and M bands are sensitive to this feature, causing a decrease of flux in the red part of L band and blue part of M band. We do not see this spectral shape in either the observed L or M band spectra. 

To put further constraints on the PH$_3$ abundance and therefore the chemistry and mixing in WISE 0855, the spectral region inaccessible from the ground due to CO$_2$ absorption from 4.2--4.5 \micron\ is needed. \emph{JWST} will be perfectly suited to this purpose, and a detection or upper limit of PH$_3$ will likely be an early result from Y dwarf observational programs. 

\subsubsection{A missing deep continuum opacity source?} \label{deepcloud}

\begin{figure}[t]
\center \includegraphics[width=3.5in]{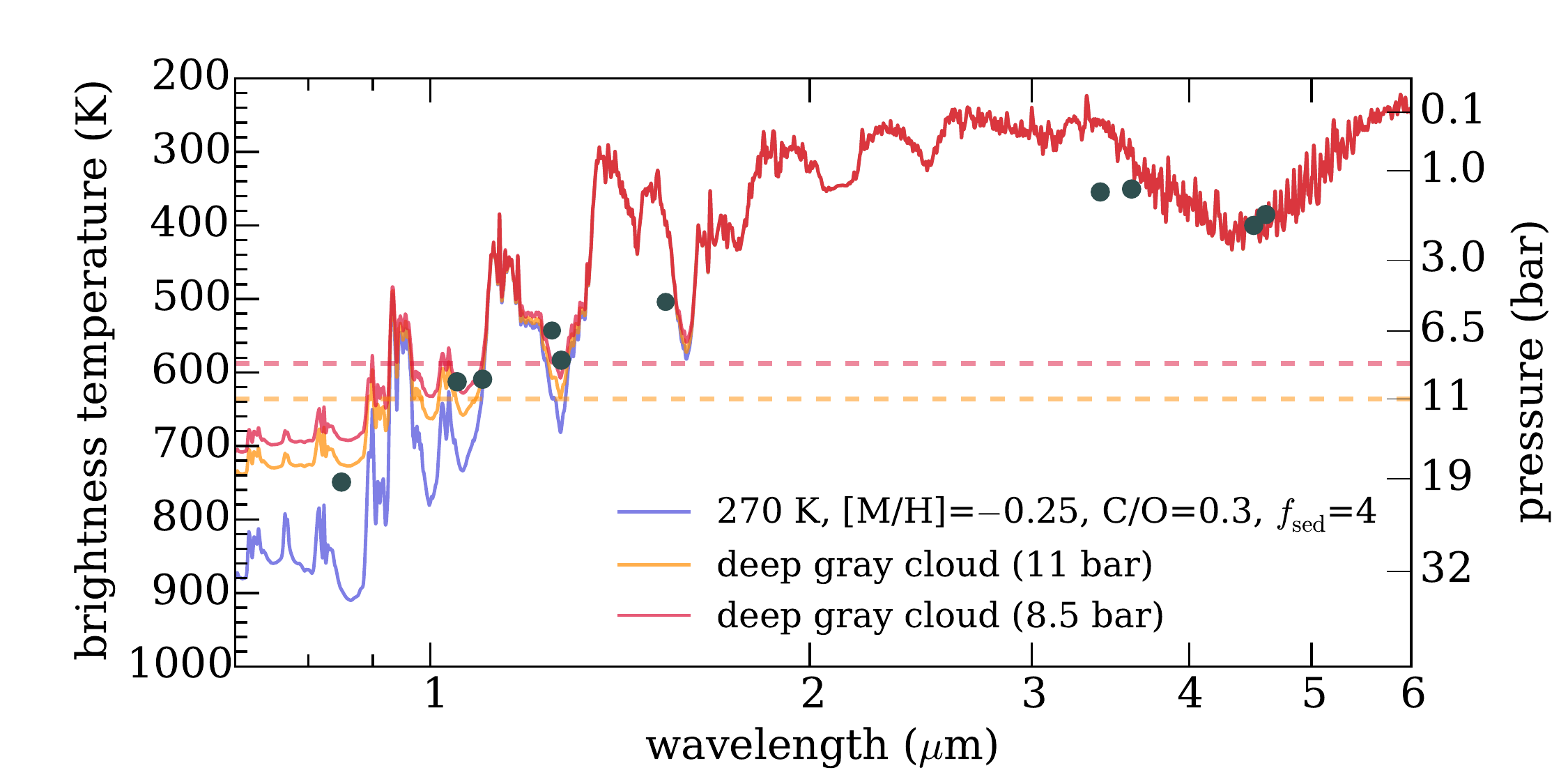}
\vspace{-11mm}
\center \includegraphics[width=3.5in]{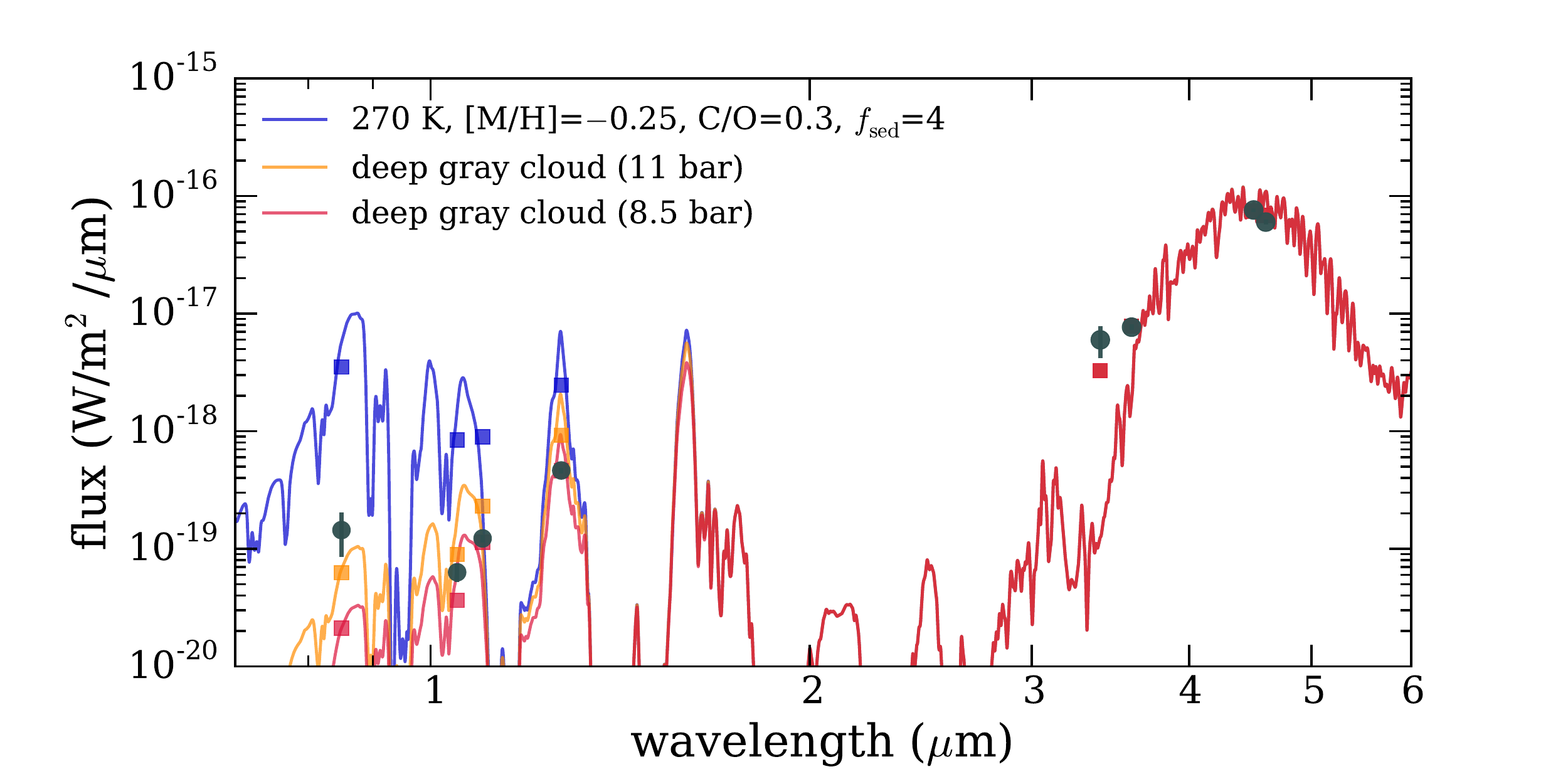}
 \caption{Model spectra including the effect of a deep gray continuum opacity source. The top panel shows the brightness temperature of model spectra as solid lines and the data as solid points. Dashed lines show the locations of $\tau=1$ gray cloud introduced into the calculation of the spectra. The bottom panel shows model spectra as solid lines, model synthetic photometry as filled squares, and data as points with error bars. The models including the deep gray cloud have reduced near-infrared flux more consistent with the observed photometry.  }
\label{deepcloud_specs}
\end{figure}

\begin{figure}[h]
\center \includegraphics[width=3.3in]{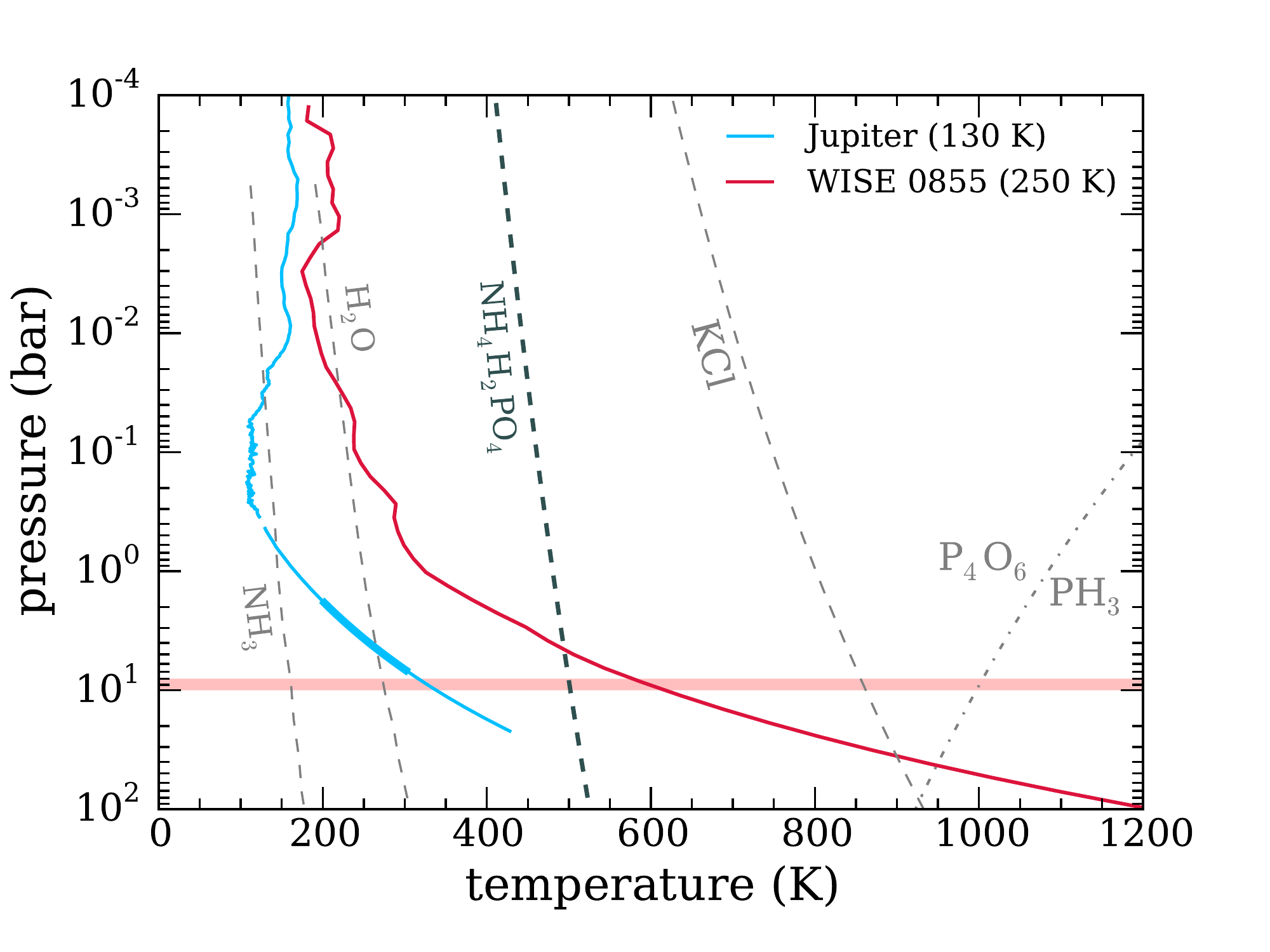}
 \caption{Pressure-temperature profile of WISE 0855 compared to Jupiter. The solid lines show model pressure-temperature profiles; the WISE 0855 profile has \teff=270 K, \fsed=4, [M/H]=$-$0.25, and C/O=0.3. The dashed gray lines represent condensation curves for each species, where the partial pressure of gas (assuming solar composition) is equal to the saturation vapor pressure. The dash-dot gray line shows the phosphorous chemistry, where the atmosphere transitions from PH$_3$-rich to P$_4$O$_6$-rich in chemical equilibrium. The red bar shows the approximate location of the continuum opacity source we propose here. The NH$_4$H$_2$PO$_4$ condensation curve is calculated assuming condensation from PH$_3$ (PH$_{3}\approx \Sigma$P) following the work of \citet{Fegley94}; cf. \citet{Visscher06}. }
\label{deepcloud_ptprof}
\end{figure}

None of the models, across a range of metallicities, cloud, C/O ratios, and temperatures, fit both the near- and mid-infrared photometry simultaneously. In most cases, models that fit the mid-infrared photometry and spectra are much brighter than the observations in the near-infrared. Because there is very little near-infrared gas phase opacity in a cold object (because a number of molecular species have condensed at cold temperatures), the model spectra probe deep layers in the near-infrared (700--900 K, or 20--40 bar). A continuum opacity source at a temperature around 600 K (around 8--10 bar) would decrease the flux substantially at those short wavelengths without changing the mid-infrared spectrum at all. 

Models illustrating the effect of such a continuum opacity source are shown in Figure \ref{deepcloud_specs}. A standard model (with a water ice cloud) is shown in blue. The same pressure--temperature profile and abundance profile is used to generate two new spectra, including a deep cloud with $\tau=1$, $\omega$=0, and asymmetry parameter $g_0$=0 in each layer, extending to pressures of 11 and 8.5 bar respectively. The models including this deep cloud layer have identical mid-infrared flux to the original model, but substantially lower thermal emission at near-infrared wavelengths, matching the observed near-infrared flux much more closely. 

We note that in Figure \ref{waterclouds_specs}, models with clouds at the altitudes of water ice clouds, but with more absorptive optical properties, also have substantially lower near-infrared thermal emission compared to the water ice clouds. It is possible that some of the assumptions made about the clouds, including that they are spherical homogeneous particles, are not applicable here. Different shaped water ice particles, or inhomogeneous ice particles, could increase the single scattering albedo at some wavelengths. Further work should be done to investigate the formation of water ice clouds in microphysical cloud models \citep[e.g.,][]{Helling06, Helling08b}

If this additional opacity source is not some form of ``dirty" water ice, it is possible that another cloud is condensing around 500--700 K. One possibility is that the phosphorous is condensing into a condensate such as ammonium dihydrogen phosphate (NH$_4$H$_2$PO$_4$). We show the condensation curve for this species compared to P--T profiles for WISE 0855 and Jupiter in Figure \ref{deepcloud_ptprof}, along with the approximate location of the gray opacity source in Figure \ref{deepcloud_specs}. This material has been predicted to condense in cool planetary and brown dwarf atmospheres \citep{Fegley94, Visscher06}, though it does not form a cloud in Jupiter's atmosphere because of the disequilibrium PH$_3$ chemistry. Nonetheless as the phosphorous chemistry in WISE 0855 appears to be distinct from Jupiter's (since we do not see PH$_3$ features in the spectrum) it is a candidate for the additional deep opacity source we propose here. If NH$_4$H$_2$PO$_4$ condenses, it would remove phosphorous from the gas phase chemistry, which could also explain lack of PH$_3$ features in the spectrum.

\section{Discussion}

\subsection{What Will \emph{JWST} Reveal About WISE 0855?}

WISE 0855 is a prime candidate for detailed characterization with \emph{JWST}, and as such it is being observed in a number of Guaranteed Time Observers (GTO) programs. For example, in the first years of \emph{JWST}'s lifetime it will be observed by three of the near-infrared instruments (NIRSpec (G395 filter and PRISM mode), NIRCam, NIRISS) and the mid-infrared instrument (MIRI/MRS), and will likely be proposed for additional observations by Guest Observers (GO) programs. 

These measurements will allow us to observe the spectrum of WISE 0855 at higher resolution and higher signal-to-noise. They will likely confirm the presence of water and methane, and also detect ammonia. Trace species with features in the mid-infrared such as phosphine, deuterated methane, carbon monoxide, carbon dioxide, hydrogen cyanide, and acetylene will either be detected or have upper limits placed on their abundance. The presence of water ice clouds will likely be confirmed or refuted.

The abundances of the major species can be used to determine the metallicity and C/O/N ratios within WISE 0855's atmosphere. The trace species can be used to study vertical mixing and chemistry in the atmosphere, comparing to Jupiter's atmosphere and other warmer brown dwarfs.

\section{Conclusion}

We presented a ground-based spectrum of the coldest known brown dwarf, WISE 0855, from 3.4--4.14 \micron. This L band spectrum shows strong methane features, as predicted by atmospheric models. WISE 0855 now has a rich dataset of photometry and spectroscopy, including photometry from 0.8 to 15 \micron\ and ground-based spectra in the L and M bands. With these data and a new set of atmospheric models that include different metallicities, C/O ratios, and water ice clouds, we probe the properties of WISE 0855. 

We find that WISE 0855's mid-infrared photometry can be better matched with models that include depleted methane abundance; further modeling work is needed to determine the root cause of this apparent depletion. Using self-consistent non-gray water ice cloud models, we show that there is evidence for water ice clouds in M band, as suggested by \citet{Skemer16}. We confirm the distinct lack of PH$_3$ compared to Jupiter's atmosphere. We also see that WISE 0855's near-infrared photometry is faint compared to models that fit the mid-infrared data, and suggest that a deep continuum opacity source could readily mask the near-infrared flux. One possibility for this deep opacity source includes a cloud of ammonium dihydrogen phosphate. 

With the upcoming launch of \emph{JWST}, we will soon probe WISE 0855 and other Y dwarfs in greater detail, detecting molecules and clouds in their atmospheres. WISE 0855 bridges the temperature regime between warmer exoplanets and the solar system giant planets, allowing us to study the physics of cool planetary atmospheres. Using WISE 0855 as a benchmark ``water cloud planet'', we will develop a deeper understanding of the processes that shape these atmospheres in advance of future missions that probe a diverse range of exoplanetary systems.

\acknowledgements
This work benefited from the Exoplanet Summer Program in the Other Worlds Laboratory (OWL) at the University of California, Santa Cruz, a program funded by the Heising-Simons Foundation. This work was performed under contract with the Jet Propulsion Laboratory (JPL) funded by NASA through the Sagan Fellowship Program executed by the NASA Exoplanet Science Institute. We thank Sandy Leggett for providing L band spectra of additional brown dwarfs. Based on observations obtained via the Fast Turnaround program GN-2017A-FT-6 at the Gemini Observatory, which is operated by the Association of Universities for Research in Astronomy, Inc., under a cooperative agreement with the NSF on behalf of the Gemini partnership: the National Science Foundation (United States), the National Research Council (Canada), CONICYT (Chile), the Australian Research Council (Australia), Minist\'{e}rio da Ci\^{e}ncia, Tecnologia e Inova\c{c}\~{a}o (Brazil) and Ministerio de Ciencia, Tecnolog\'{i}a e Innovaci\'{o}n Productiva (Argentina). The authors wish to recognize and acknowledge the very significant cultural role and reverence that the summit of Maunakea has always had within the indigenous Hawaiian community. We are most fortunate to have the opportunity to conduct observations from this mountain.

%\bibliographystyle{apj}
%\bibliography{references}

\end{document}